\documentclass[prb,preprint,aps,showpacs]{revtex4} 
\usepackage{graphicx} 
\usepackage{tabularx}
\usepackage{dcolumn} 
\usepackage{color}
\usepackage{bm}
\usepackage{amsmath}
\usepackage{amssymb}

\begin{document} 
 \title{
Generic thermodynamics of condensates with extremely small superfluid density \\
---Application to UTe$_2$ and hidden second phase transition---
}
\author{Kazushige Machida} 
\affiliation{Department of Physics, Ritsumeikan University, Kusatsu 525-8577, Japan} 
\date{\today} 
\begin{abstract}
Motivated by recent experimental findings on UTe$_2$ via NMR and
specific heat experiments, we theoretically explore the thermodynamic signatures of
condensates with small superfluid density and characterize their physical properties.
We identify the hidden superconducting phase $A_2$ composed of spin-up Cooper pairs
 at lower temperature ($T$)
and field ($H$) under ambient pressure of UTe$_2$ and show that $A_2$ reappears at higher $H$ for all
three principal field directions, $a$, $b$, and $c$-axis. The phase transition lines from the $A_1$ phase  with spin down pairs
to the $A_2$ phase in the $H$-$T$ plane are all horizontal in common, implying  a universal physical origin due to the
originating from the magnetic-energy gain of the spin-up pairing state with increasing magnetic field. This multiple-phase diagram is extended to these under pressure ($P$),
resulting in a simple picture based on non-unitary spin-triplet pairing symmetry $^3B_{3u}$, which enables us to
consistently capture the evolution of the superconducting phases throughout the entire  $H$-$T$-$P$  phase space.
\end{abstract}

\maketitle

\section{Introduction}
The superfluid density $\rho_{\rm s}$ is one of the fundamental characteristics of a superconducting condensate, governing its electromagnetic response~\cite{pickett}.
A class of superconductors exhibits an extremely low superfluid density, or the superfluid phase
stiffness among strongly correlated materials, such as under- and over-doped high $T_{\rm c}$ cuprates~\cite{uemura,homes,cooper} 
or twisted multi-layered graphenes~\cite{randeria,perge} and moir\`{e} systems~\cite{mak}.
The former is characterized by the Uemura plot or the Homes law, where superconductivity (SC) is realized under low-doped carriers. 
In the latter, a Dirac flat band with low Fermi velocity $v_F$$\sim$1km/s sustains SC with the 
unusually small BCS gap ratio, namely  $\Delta/k_{\rm B}T_{\rm c}$$\sim$$0.05$ (energy gap $\Delta$ and 
$T_{\rm c}$ transition temperature) that is far below 1.76 for the weak coupling BCS formula.
Since $\rho_{\rm s}$ corresponds to the depleted normal electron number density 
estimated by $\rho_{\rm s}$$\propto$$N(0)\Delta$ in the filled part of the electron spectrum ($N(0)$ density of states at the Fermi level),
an exotic SC state with the extremely low  $\rho_{\rm s}$ may be realized.
These characteristics are shared by a heavy Fermion superconductor UTe$_2$, of interest in this paper, since 
$v_F$ is a few km/s with an extremely narrow ``flat'' band whose width is comparable to the Kondo temperature $\sim$30K
produced by the Kondo effect. In fact, Kamat, {\it et al.}~\cite{kamat} report clear evidence for vanishing phase stiffness
realized in UTe$_2$ through their ultrasound measurement.

The present work is motivated by a recent  intriguing NMR measurements by Matsumura, {\it et al.}~\cite{matsumura0}:
They measure the nuclear spin relaxation time $1/T_1$ in high-quality samples of
UTe$_2$ for $H\parallel b$-axis at lower $T$ and low $H$,
discovering the anomalous enhancement, or unexpected local peak structure in $1/T_1$ around 0.1K for $H$=0.38T, 0.65T, 
and 0.8T, whereas the Hebel-Slichter type
$1/T_1$ enhancement just below $T_{\rm c}$=2.1K is absent. The absence of the Hebel-Slichter peak is expected for spin-triplet superconductors, whereas the origin of the low-temperature enhancement remains unknown.
Intriguingly, the recent experiment~\cite{totsuka} also detects an anomalous increase of the specific heat at lowest $T$ around 0.3K
in the absence of $H$ just before an unknown specific heat uprise toward $T$$\rightarrow0$.

The objectives of this paper are threefold:
 To understand these seemingly independent experimental reports~\cite{matsumura0,totsuka} under the context of a spin-triplet state with the extremely small superfluid density 
by investigating generic physical properties of such a condensate,
to identify  the presence of the ``hidden'' second phase transition $T_{\rm c2}$=0.1K$\sim$0.3K at both $H$=0 and low $H$,
and to explain why it is hidden. 
We will try to draw a unified picture for multiple phase diagrams in the $H$-$T$ plane under ambient and pressure settings 
by combining the previously known high field second phase, so-called SC2
for $H$$\parallel$$b$-axis, which is the target condensate of the present  paper.

\begin{figure}
\begin{center}
\includegraphics[width=15cm]{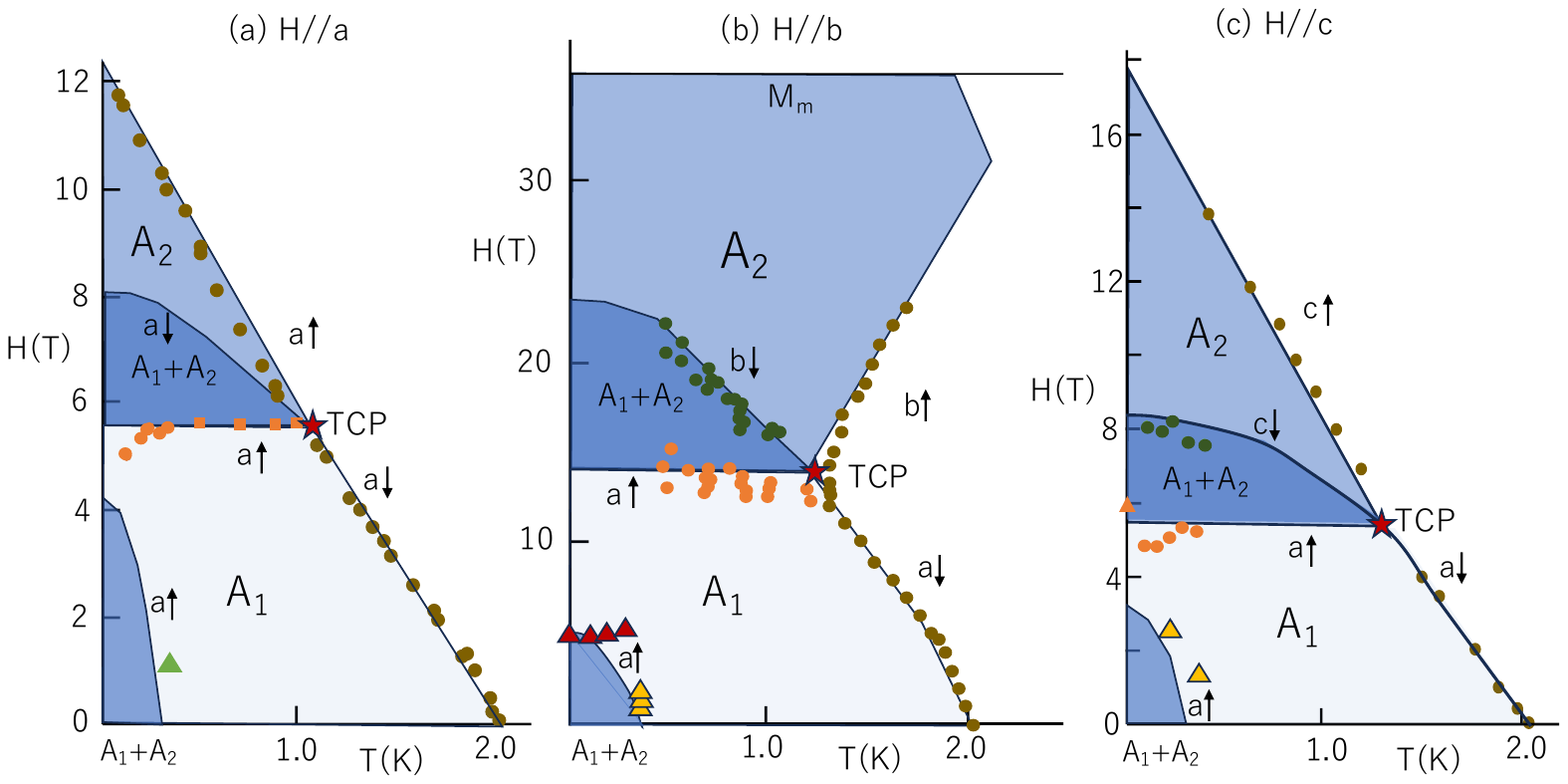}
\end{center}
\caption{Phase diagrams consisting of $A_1$, $A_2$, and its mixture 
$A_1$+$A_2$ for (a) $H$$\parallel$$a$-axis with the data~\cite{tokiwa0} for $H_{{\rm c2},a}$, and other data
points~\cite{totsuka, sangyun}.
(b) $b$-axis where the data points come from \cite{sakai,aoki,rosuel,matsumura0,roman,sangyun}, and (c) $c$-axis
where the data points come from \cite{matsumura0,matsumurachi,roman,sangyun}.
 We indicate the spin quantization axis and its polarization direction for
each phase. TCP denotes the tetracritical point with the red star symbols.
The detailed explanations of each phase diagram are given in Section IV.
\label{pd-abc}}
\end{figure}

Before embarking on the detailed investigations, it is necessary to summarize the present status of the problem
associated with the heavy Fermion superconductor UTe$_2$ from our point of view (see also general reviews~\cite{review,review1}).
To construct a unified picture of the UTe$_2$ problem, we notice the following facts:

\noindent
(1) As shown in Fig.~\ref{pd-abc}(b) 
under the ambient pressure $P$=0, the $H$-$T$ phase diagram for $H$$\parallel$$b$-axis consists of at least two phases, SC1
denoted as A$_1$ in low $H$  and SC2 (A$_2$) in high $H$ divided by a horizontal internal line at $H$=14T~\cite{sakai}. 
The double transition between SC1 (A$_2$) and SC2 (A$_2$)
 is observed by the specific heat experiment~\cite{rosuel}, constituting a tetracritical point where the four second order phase
 transition lines meet. This situation is thermodynamically allowed and thus stable.

 \noindent
(2) Under pressure $P$, $T_{\rm c}$ goes down continuously while at $P_{\rm cr}$=0.2GPa the second transition 
$T_{\rm c2}$ appears and goes up, reaching a maximum $T_{\rm c2}$$\sim$3K at around $P$=0.7GPa.
Thus, the three second order transition lines meet at $P$=0.2GPa, which is thermodynamically unstable,
suggesting a missing second order line between $P$=0 and $P$$_{\rm cr}$=0.2GPa (see later in Fig.~\ref{pd-p} for details).
If $T_{\rm c2}$ is found to be nonzero at $P$=0 and continues to $P$=0.2GPa, $P$$_{\rm cr}$ becomes a tetracritical point.
Then, the $P$-$T$ phase diagram becomes thermodynamically stable.

 \noindent
 (3) The high field phase SC2 (A$_2$) exhibits an extremely small jump compared with that of SC1 (A$_1$), which is conventional,
 in the specific heat experiment~\cite{rosuel}.
Note that under $P$ the phase stiffness of the high (low) $T$ phase is extremely small (conventional)~\cite{kamat},
indicating the close connection of SC2 (A$_2$) in $P$=0 and the high $T$ phase above $P$=0.2GPa.
This internal relationship is clarified in this paper.

 \noindent
 (4) According to Tokiwa, {\it et al.}~\cite{tokiwa0}, in the ambient $P$=0 
 the horizontal transition line at around $H$=5$\sim$6T divides low  and high $H$ phases
 for $H$$\parallel$$a$-axis, similar to that for the $b$-axis above, suggesting SC1 (A$_1$) and SC2 (A$_2$) respectively
 as shown in Fig.~\ref{pd-abc}(a).
 In this paper, we demonstrate that this is indeed the case. 
 The phase diagram for $H$$\parallel$$c$-axis shown in Fig.~\ref{pd-abc}(c) is also a similar internal structure
 subdivided by SC1 (A$_1$) and SC2 (A$_2$) at around $H$=6T where $H_{\rm c2}$ has a kink and the Knight shift (KS) 
 saturates to the
 normal-state value~\cite{matsumurachi}, above which KS stops decreasing at $T_{\rm c}$.

\noindent
 (5) We emphasize a universal feature that all the phase diagrams for $H$$\parallel$$a$, $b$, and $c$-axes are
 subdivided into the two phases $A_1$ and $A_2$ as displayed in Fig.~\ref{pd-abc}.
 One of the main purposes of this paper is to derive these phase diagrams.
 By measuring magnetization curves for all directions, Shimizu, {\it et al.}~\cite{shimizu} have recently found 
 the anomalies associated with the internal phase transition lines for $H$$\parallel$$a$-axis.

 \noindent
 (6) As for the KS experiments~\cite{matsumura,ishida1,ishida2,ishida3,ishida4,ishida5,kinjo,kitagawa} 
 that probe the Cooper pair spin structure,
 the spin susceptibilities for three major directions for the $a$, $b$, and  $c$-axes decrease below $T_{\rm c}$.
 We have argued that this can be understood as the Cooper-pair spins are aligned antiparallel
 to the $a$-axis, or the d-vector contains the components along the $b$-axis and $c$-axis 
 in our series of papers~\cite{machida1,machida2,machida3,machida4,machida5,machida6,tsutsumi,machida7},
 indicating that the spin part of the order parameter (OP) can be written as ${\hat{\bold b}}\pm i{\hat{\bold c}}$
 with ${\hat{\bf b}}$ and ${\hat{\bold c}}$ unit vectors, namely the equal spin state consisting of $\uparrow\uparrow$
 or $\downarrow\downarrow$ spin pairs along the $a$-axis in the absence of $H$.
 Thus OP is non-unitary~\cite{machida0,annett,ozaki1,ozaki2,ramires}.

 \noindent
 (7) The orbital part of OP, or the gap structure is characterized by line nodes running along the $b$-axis
 on the $\beta$ Fermi surface (FS), namely described by $\sin(ak_a/2)$ with $a$ being the lattice constant along the $a$-axis.
 It should be noted that this gap function leaves the full gap for the $\alpha$ Fermi surface.
 Both $\alpha$ and $\beta$ Fermi surfaces have rectangular cross-sections open along the $c$-axis  and undulate 
 (the location of the line nodes in the reciprocal space, see Fig. 4 in Ref.~[\onlinecite{totsuka}]).
 Therefore, the overall OP is $({\hat{\bold b}}\pm i{\hat{\bold c}})\sin(ak_a/2)$, belonging to ${^3}$${\rm B}_{\rm 3u}$
 irreducible representation in the weak spin-orbit coupling classification scheme~\cite{machida0,annett,ozaki1,ozaki2}.
The coexistence of the line node $\beta$-FS and full gap $\alpha$-FS turns out to be crucial to interpret the $1/T_1$ data 
by Matsumura, {\it et al.}~\cite{matsumura0} as we will see later. 
We also note that this particular pairing function is compatible with the ultrasound measurements
by Theuss, {\it et al.}~\cite{ramshaw}, proving that the orbital OP component is single. 

 \noindent
 (8) The phase diagram in the $P$-$T$ space is now being established experimentally.
 In particular, the recent report~\cite{kamat2} shows that the tetracritical point is situated at $P$=0.19GPa
 where the four second order phase transition lines meet. The newly discovered fourth 
 line with a negative slope as a function of $P$ will be critically discussed in the main text.

In the following, we use the notations $A_1$ and $A_2$ in pairs to distinguish the two phases with the equal spin pairs
which contain the information on the spin direction, or the spin quantization axis and either $\uparrow$$\uparrow$ or 
$\downarrow$$\downarrow$ spin pairs.
We denote the two superconducting transition temperatures by $T_{\rm c1}$ and $T_{\rm c2}$ to merely distinguish the
two transitions under a given circumstance such as magnetic field $H$ or pressure $P$.

\section{Theoretical considerations}

It is customary to introduce the adjustable gap ratio 
$\alpha=\Delta_0/k_{\rm B}T_{\rm c}$ with $\Delta_0=\Delta(T=0)$
when analyzing $T_1$  or specific heat jump at $T_{\rm c}$ in superconductors to identify the gap structure. 
For example, the sharp decrease of $1/T_1(T)$ just below  $T_{\rm c}$ without the Hebel-Slichter enhancement
for d-wave cuprate superconductors or heavy Fermion superconductors such as
CeCu$_2$Si$_2$~\cite{cecu2si2} and UBe$_{13}$~\cite{maclaughlin} is explained by taking $\alpha$ as large as $\alpha$=4.0~\cite{kitaoka}.
Since the strong coupling effects due to the electron-phonon interaction are known to give rise to an arbitrarily
large  $\alpha$ parameter~\cite{mitrovic}, such a large gap ratio is not unexpected.
In UTe$_2$ Nakamine, {\it et al.}~\cite{ishida1} employed $\alpha$=3.5 (full gap), 3.8 (point node),
 and 5.9 (line node), which depend on the assumed nodal structures, to fit  their $1/T_1(T)$ data near $T_{\rm c}$.
The specific heat jump $\Delta C/\gamma_{\rm N}T_{\rm c}$=1.43 below $T_{\rm c}$ ($\gamma_{\rm N}$ 
Sommerfeld coefficient in the 
normal-state) given by the BCS theory often exceeds this value, which is explained in terms of the
strong coupling effects above. 
For UTe$_2$ the observed $\Delta C/\gamma_{\rm N}T_{\rm c}\sim$2.5~\cite{totsuka} for a high quality sample 
is understood by taking $\alpha$=2.5.

In contrast to the conventional strong-coupling scenario with $\alpha>1.76$, we focus on the opposite limit,
corresponding to a condensate with the extremely small 
superfluid density or lower phase stiffness.

\subsection{Nuclear relaxation time T$_1$}

We begin with the nuclear spin-lattice relaxation rate for $T_1$ normalized by its normal-state $T_{\rm 1N}$ 
for a spin-triplet pairing~\cite{sigrist}:

\begin{eqnarray}
{T_{\rm 1N}\over T_1}={2\over N(0)^2}\int_0^{\infty}dE N(E)N(E+\hbar\omega_0)\Big(-{df(E)\over dE}\Big)
\label{t10}
\end{eqnarray}

\noindent
where $N(E)$ is the quasiparticle density of states (DOS), $\hbar\omega_0$ the NMR resonance frequency,
and $f(E)$ the Fermi-Dirac distribution function. 
The contribution coming from the so-called anomalous DOS, corresponding to the coherent factor vanishes
because the unconventional SC is characterized by $\langle \Delta(k)\rangle_{\rm FS}$=0 
in general ($\langle\cdots\rangle_{\rm FS}$ denotes the average over the Fermi surface).

\subsubsection{Full gap case}

We first consider the fully gapped case.

\begin{eqnarray}
{T_{\rm 1N}\over T_1}=2\int_{\Delta(T)}^{\infty}dE{E^2\over {E^2-\Delta^2(T)+i\eta}}\Big(-{df(E)\over dE}\Big)
\label{t1form}
\end{eqnarray}

\noindent
with the infinitesimally small non-dimensional constant $\eta$ or the smearing parameter of the order of $\eta=\hbar\omega_0/\Delta_0$
to regularize the gap-edge singularity. Equation~(\ref{t1form}) is valid for zero field or lower fields.

\begin{figure}
\begin{center}
\includegraphics[width=12cm]{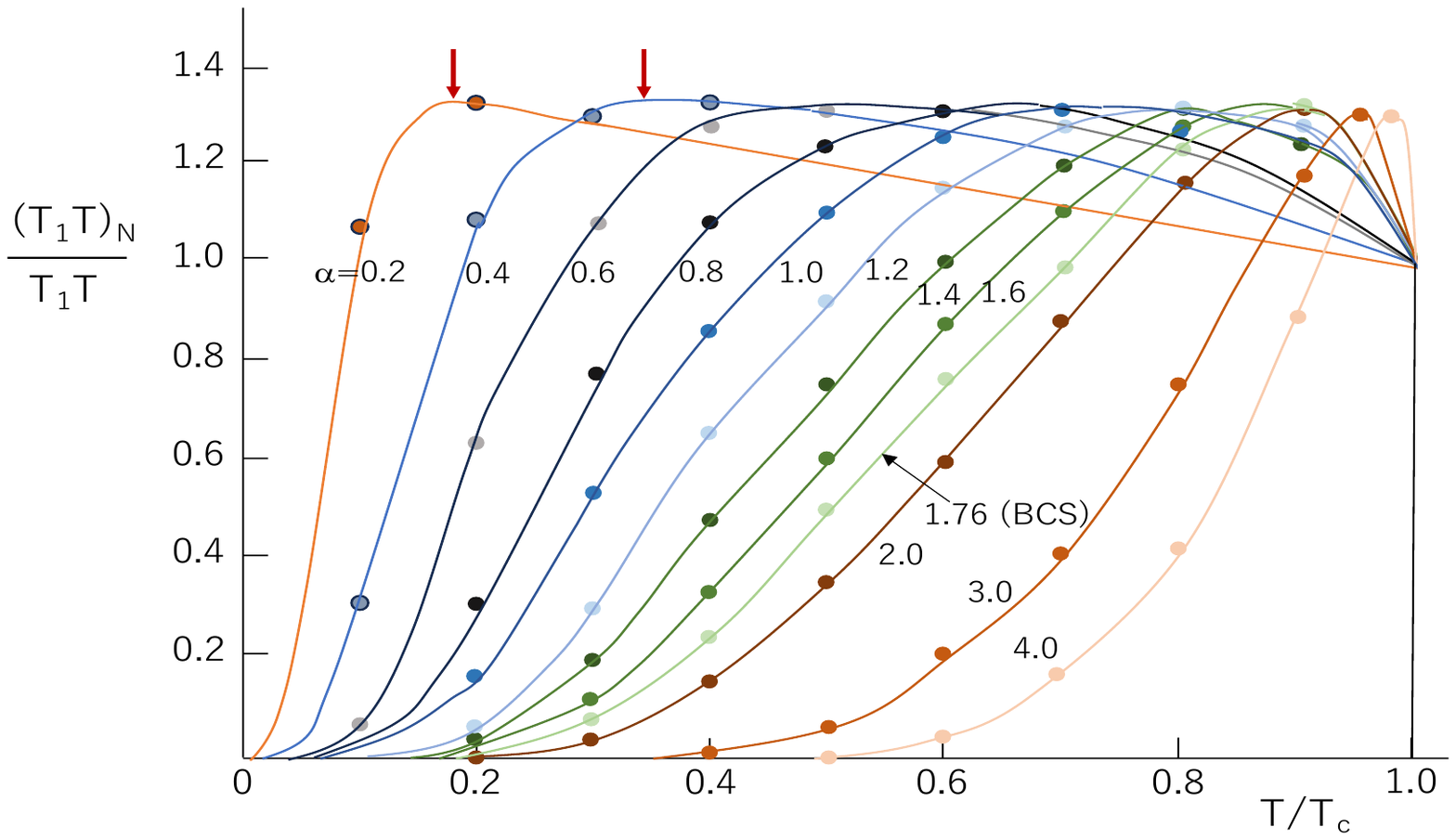}
\end{center}
\caption
{The normalized $(T_1T)_{\rm N}/(T_1T)$ as a function of $T$ for various $\alpha$ 
values from larger to smaller ones,
including the BCS case with the smearing parameter $\eta$=0.01. The downward arrows indicate the estimated peak positions  for $\alpha$=0.4 and
0.2, which agree well with the analytical estimate given by Eq.(\ref{t1max}).
\label{t1bcs}}
\end{figure}

As shown in Fig.~\ref{t1bcs},
we calculate $(T_1T)_{\rm N}/(T_1T)$ normalized by the normal-state value $(T_1T)_{\rm N}$ for various $\alpha$ values
for the full gap case.
Figure~\ref{t1bcs} shows that 
for $\alpha$ larger than the BCS value 1.76, the Hebel-Slichter peak occurs just below $T_{\rm c}$
and $(T_1T)_{\rm N}/(T_1T)$ is followed by an exponential suppression.
As $\alpha$ decreases from the BCS value, the Hebel-Slichter peaks shift to lower $T$ while keeping their height.
The characteristic exponentially decreasing $T$ region is shifted toward lower temperatures.

The peak position can be estimated analytically.
By rewriting Eq.~(\ref{t1form}) as

\begin{eqnarray}
{(T_1T)_{\rm N}\over T_1T}=\int_{0}^{\infty}dx{x^2\over {x^2-{\alpha^2\over 4}({\delta\over t})^2+i\eta}}\cdot{dx\over \cosh^2(x)}
\label{ct1}
\end{eqnarray}

\noindent
with $t=T/T_{\rm c}$, and $\delta=\Delta(T)/\Delta_0$, 
the maximum occurs when ${\alpha^2\over 4}({\delta\over t})^2$$\sim$0.3 for smaller $\alpha$.
It implies that

\begin{eqnarray}
{T_{\rm max}\over T_{\rm c}}={\alpha \over 2\sqrt{0.3}}=0.9\alpha.
\label{t1max}
\end{eqnarray}

\noindent
The maximum positions are indicated for $\alpha$=0.2 and 0.4 in Fig.~\ref{t1bcs} by the down arrows.
It is seen from Fig.~\ref{t1bcs} for $\alpha$=0.2 that the Hebel-Slichter peak shifts to lower $T$ around $T=0.1T_{\rm c}$
with the peak height $(T_1T)_{\rm N}/(T_1T)_{\rm max}$$\sim$1.3.

\begin{figure}
\begin{center}
\includegraphics[width=12cm]{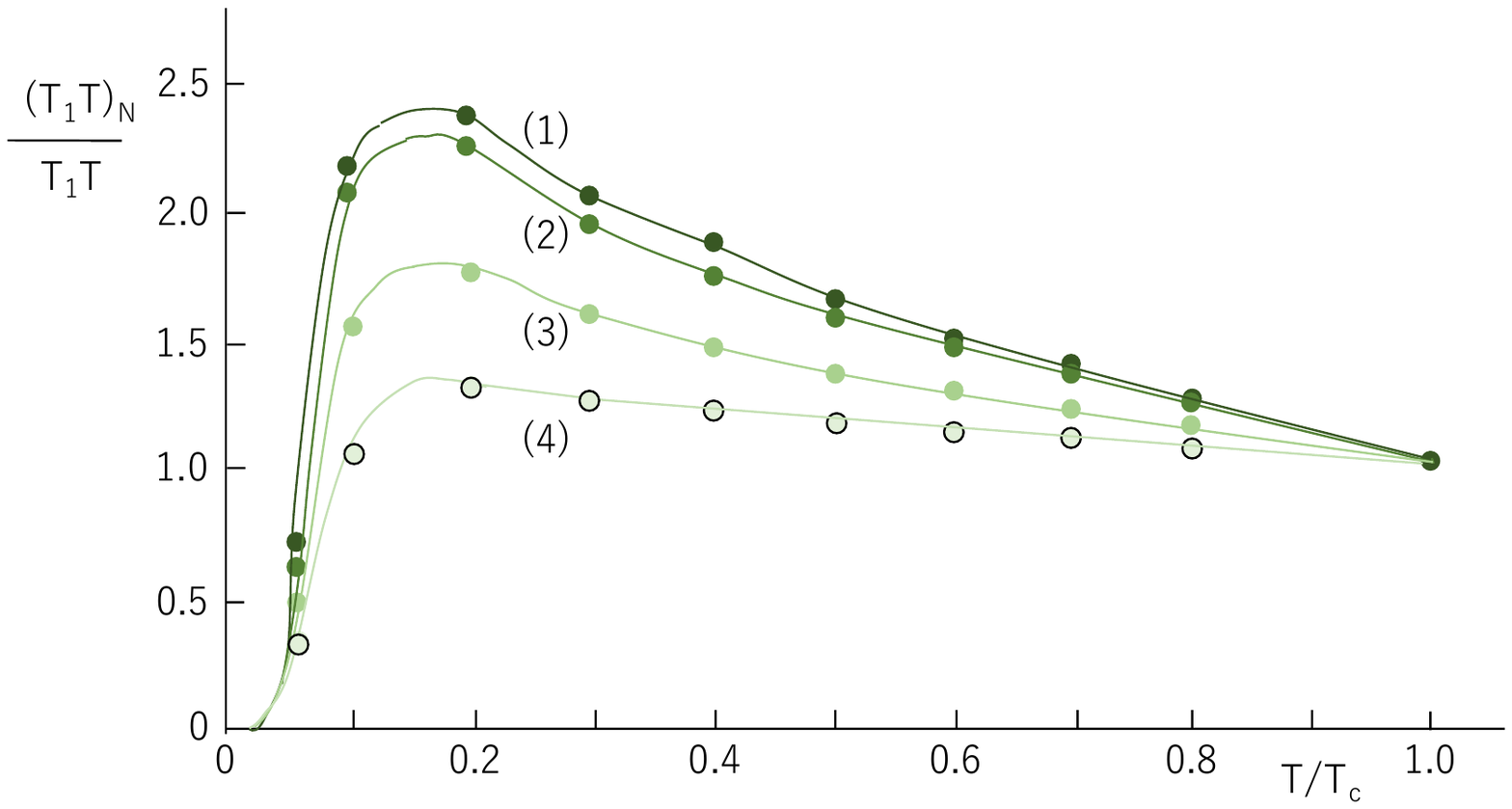}
\end{center}
\caption
{$(T_1T)_{\rm N}/(T_1T)$ as a function of $T$ for various $\eta$. (1) $\eta$=0.00005, (2) 0.0001, (3) 0.001,
and (4) 0.01. The peak heights progressively increase as $\eta$ becomes small.
\label{t1eta}}
\end{figure}

Next, we examine the peak height dependence on the smearing parameter $\eta$.
For $\alpha$=0.2, $(T_1T)_{\rm N}/(T_1T)$ is calculated for various  $\eta$ values
in Fig.~\ref{t1eta}.
Figure~\ref{t1eta} shows that as $\eta$ decreases, the Hebel-Slichter peak exceeds twice the normal-state value compared with the corresponding normal value.

Note that the non-dimensional smearing parameter could be an order of $\sim O(10^{-6})$ for the resonance frequency MHz region and $T_{\rm c}$=2.1K
in UTe$_2$ under the clean limit situation~\cite{curro}.
Therefore, a smaller value of $\alpha$ effectively shifts the Hebel--Slichter peak to temperatures well below $T_{\rm c}$.
This observation is crucial for interpreting the experimental data of Matsumura, {\it et al.}~\cite{matsumura0}, in other words
the peak of $1/T_1$ does not occur just below $T_{\rm c}$ in the smaller $\alpha$ cases.
This is one of the characteristic signatures of superconductors with an extremely small superfluid density.

\subsubsection{Line-node and finite-field cases}

For the line node case, we can derive the formulae by choosing 
$N(E)$~\cite{sigrist} as

\begin{align}
N(E)&={\pi \over 2}{|E|\over \Delta}                             &|E|<\Delta \nonumber \\
        &={|E|\over\Delta}\arcsin({\Delta\over |E|})           &|E|>\Delta.
\label{line}
\end{align}

\noindent
Since in line node case the edge singularities at $E=\pm \Delta$ are weaker than that in the fully gapped case,
no smearing parameter is needed in the numerical computations.

\begin{figure}
\begin{center}
\includegraphics[width=12cm]{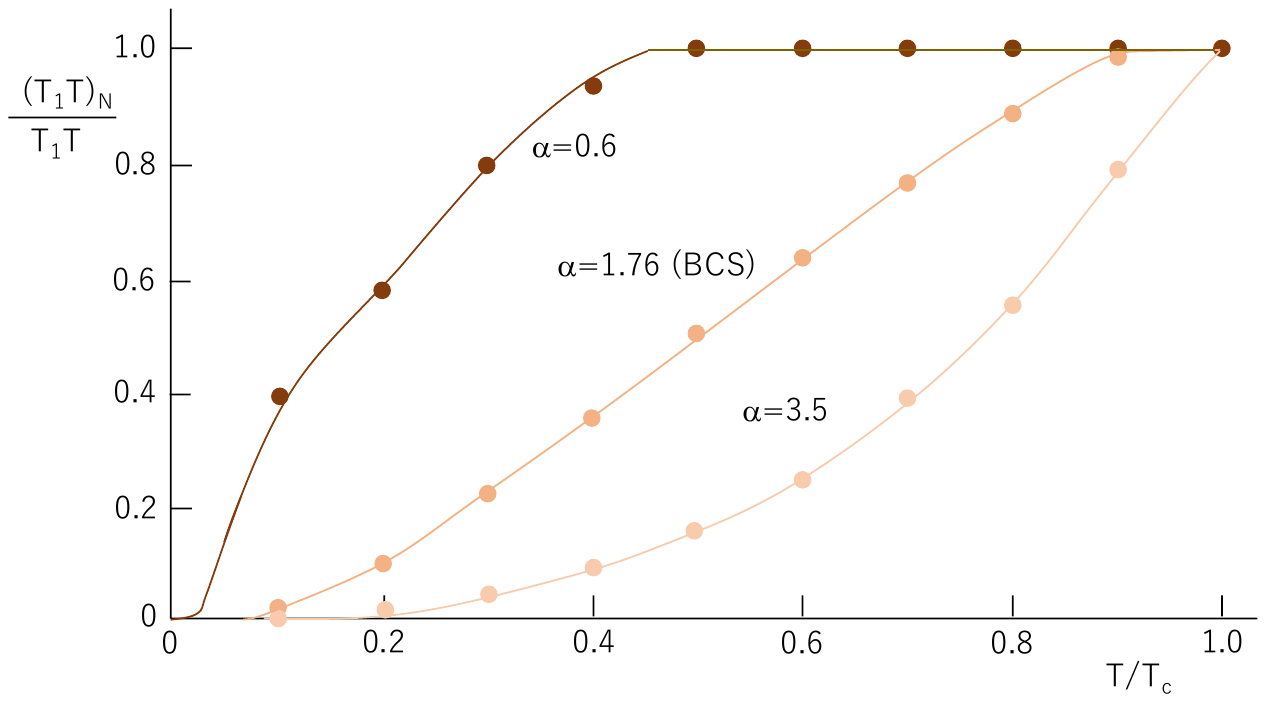}
\end{center}
\caption{$(T_1T)_{\rm N}/(T_1T)$ as a function of $T$ for various $\alpha$, including the BCS case
for the line nodes.
The sharp drop at $T_{\rm c}$ is seen for $\alpha$=3.5 while the drop is delayed to a lower $T$ for smaller $\alpha$.
\label{linenode}
}
\end{figure}

As shown in Fig.~\ref{linenode}, $(T_1T)_{\rm N}/(T_1T)$ exhibits no prominent peak
near $T_{\rm c}$ for the BCS case ($\alpha$=1.76) because the DOS singularity at the energy gap is weak.
At lower $T$ the characteristic power law behaviors $\propto T^2$ associated with the line node structure are seen.
As $\alpha$ increases to larger values, a strong decrease in $1/T_1$ is seen from it.
This confirms previous reports observed in cuprates, and heavy Fermion materials~\cite{cecu2si2,maclaughlin,kitaoka}.
On the contrary, the smaller $\alpha$=0.6 in Fig.~\ref{linenode} shows a flat $T$ region near  $T_{\rm c}$
because in this case the gap opening is delayed to a lower $T$ and the near  $T_{\rm c}$
situation persists, thus the $T^2$ power law behavior is seen at an extremely lower $T$.

Under a finite magnetic field, namely in a vortex state it is proven~\cite{nakai} that the
averaged DOS $N(E)$ over a vortex lattice unit cell at the lower energy is given universally by the linear $|E|$ near the Fermi level
on top of $\gamma(H)$ normalized by the normal-state value $\gamma_{\rm N}$ at $E$=0, namely, 

\begin{eqnarray}
{N(E)= \gamma(H)+c{ |E|\over \Delta}}
\label{dos}
\end{eqnarray}

\noindent
with $c$ being a constant given by $c=\pi/2-\gamma(H)$.
This V-shape DOS plus $\gamma(H)$ is independent of the underlying gap structures, either
the full gap, point nodes, or line nodes.

\begin{figure}
\begin{center}
\includegraphics[width=12cm]{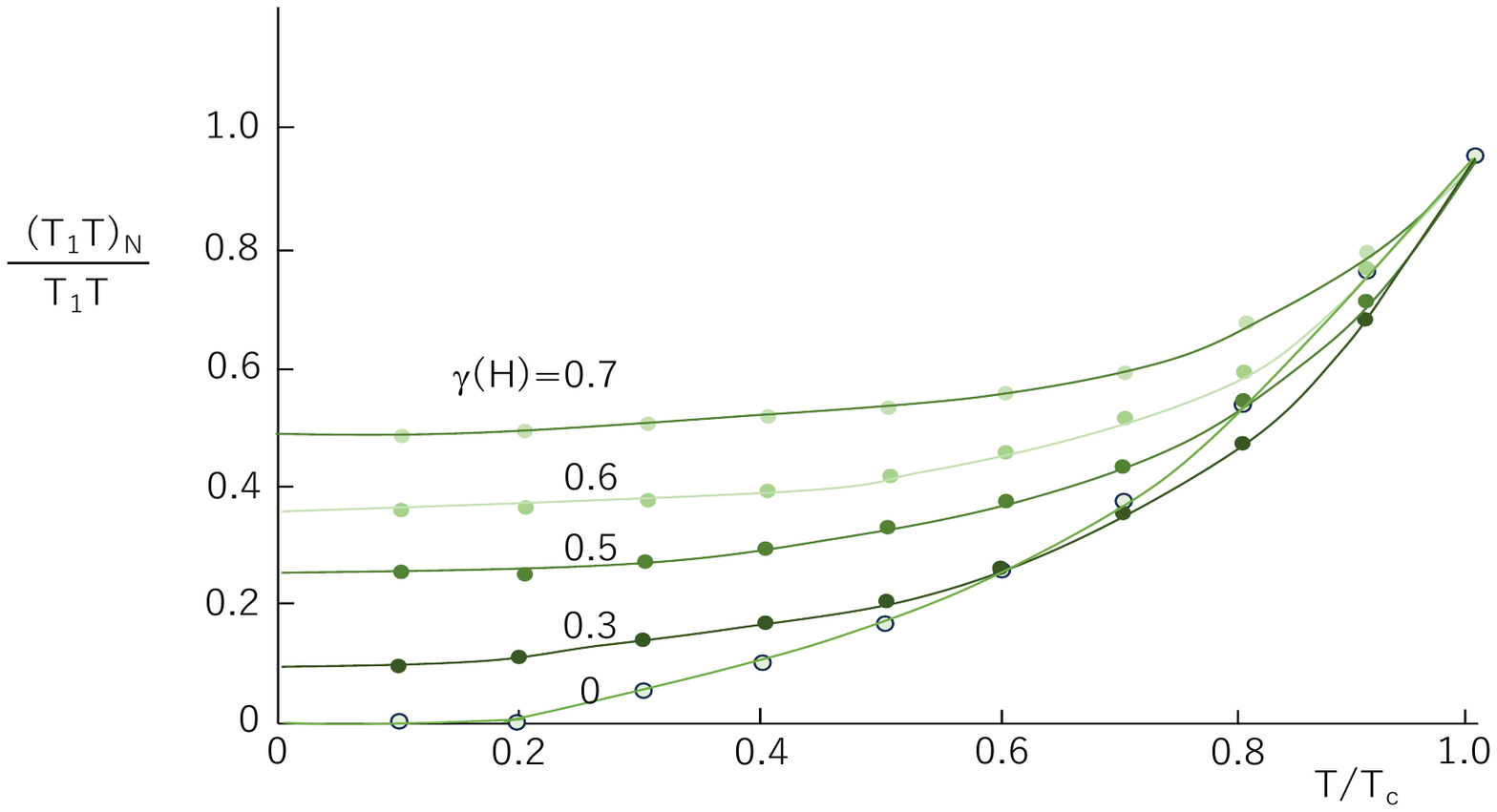}
\end{center}
\caption{$(T_1T)_{\rm N}/(T_1T)$ for the line node case as a function of $T$ in various $\gamma(H)$ ($\alpha$=3.5).
After sharp drop below $T_{\rm c}$, $(T_1T)_{\rm N}/(T_1T)$ exhibits a plateau.
The limiting values of the plateaus toward $T\rightarrow$0 are given by $\gamma^2(H)$ in Eq. (\ref{tending}).
\label{line2}
}
\end{figure}

As shown in Fg.~\ref{line2} where we evaluate $(T_1T)_{\rm N}/(T_1T)$ by assuming
 the DOS given by Eq.~(\ref{dos}),
 we find that for finite $H$ cases $(T_1T)_{\rm N}/(T_1T)$ exhibits a plateau at lower $T$,
 and that the tending limit of $(T_1T)_{\rm N}/(T_1T)$ toward $T$$\rightarrow$0 becomes $\gamma^2(H)$
 instead of $(T_1T)_{\rm N}/(T_1T)$$\rightarrow$0 in zero-field case, while the behavior near $T_{\rm c}$ remains essentially unchanged.
These results are quite reasonable since

\begin{eqnarray}
{{(T_1T)_{\rm N}\over(T_1T)}\propto N(0)^2=\gamma^2(H)}
\label{tending}
\end{eqnarray}

\noindent
at $T$=0 while near $T_{\rm c}$
DOS near the gap edge $E\sim\Delta_0$ dominates the $T$ behavior.

Before finishing the discussions on $1/T_1$, we remark the case where two condensates 1 and 2 coexist
with normal-state DOS contributions are $\gamma_1$ and $\gamma_2$.
Since the relaxation rate is additive, $1/T_1$ is written as a sum of the two condensates,
namely, $1/T_1=1/T^{(1)}_1+1/T^{(2)}_1$.
It is clear that the relaxation processes are dominated by the shorter $T_1$ or quicker process.
Thus the condensate with the longer relaxation time manifests itself only in the lower $T$ region where the first process becomes negligible.

\subsection{Specific heat}

To further characterize superconductors with an extremely low superfluid density, the specific heat provides one of the most informative thermodynamic probes.
For simplicity, we consider only the fully gapped case.
The essential conclusions, however, are equally applicable to other nodal structures.
We start with the standard formula for $C(T)$ given by

\begin{eqnarray}
C(T)=C_H(T)+C_L(T),
\label{ct}
\end{eqnarray}

\begin{eqnarray}
C_H(T)=\int_{-E_F}^{\infty}dEN(E)\big(-{\partial\Delta^2\over \partial T}\big)\big(-{\partial f\over \partial E}\big),
\label{ch}
\end{eqnarray}

\begin{eqnarray}
C_L(T)={2\over T}\int_{-E_F}^{\infty}dEN(E)E^2\big(-{\partial f\over \partial E}\big).
\label{cl}
\end{eqnarray}

\noindent
The first term dominates the behavior near ${T_{\rm c}}$ 
whereas the second governs the low-temperature limit.
From Eq.~(\ref{ch}) one readily obtains the expression of the specific heat jump at $T_{\rm c}$,
that is, by starting from Eq.~(\ref{ch}) we arrive at  
$\Delta C=N(0){\big(}-{\partial\Delta^2\over \partial T}{\big)}_{T_{\rm c}}$.
The normal specific heat $C_{\rm N}={2\pi^2\over 3}N(0)T=\gamma_{\rm N}T$.
Since $(-{\partial\Delta^2\over \partial T})_{T_{\rm c}}=3.03{\Delta_0^2\over T_{\rm c}}$,
we obtain 

\begin{eqnarray}
{\Delta C\over \gamma_{\rm N}T_{\rm c}}=0.46\alpha^2.
\label{jump}
\end{eqnarray}

\noindent
We can check that when $\alpha=1.76$, it reduces to the BCS value 1.43.
Therefore, superconductors with smaller values of $\alpha$ exhibit
a smaller jump at $T_{\rm c}$, making it harder to detect.

In Fig.~\ref{ct} we display the results for $C(T)/\gamma_{\rm N}T$ for various $\alpha$ values, including
the BCS case. Figure ~\ref{ct} shows that as $\alpha$ decreases, the specific heat jump $\Delta C(T)/\gamma_{\rm N}T$ at $T_{\rm c}$ 
decreases rapidly as described by Eq.~(\ref{jump}).
The peak shifts to lower temperatures and the peak height is also lower.
We can derive the approximate temperature $T_{\rm max}$, at which $C(T)/\gamma_{\rm N}T$ takes a maximum for small $\alpha$:

\begin{eqnarray}
{T_{\rm max}\over T_{\rm c}}={\alpha\over 2\sqrt{0.75}}=0.578\alpha
\label{max}
\end{eqnarray}

\noindent
and its height $C(t_{\rm max})/\gamma_{\rm N}T=1.17$ with $t_{\rm max}=T_{\rm max}/T_{\rm c}$.
The down arrows in Fig.~\ref{ct} show the positions of $t_{\rm max}$ for $\alpha$=0.2 and 0.4.
The peak position does not correspond to the transition temperature anymore.
The characteristic exponentially activated region is shifted toward lower temperatures and becomes progressively narrower.

An important consequence is the entropy balance:
the areas above and below  the $C(T_{\rm c})/\gamma_{\rm N}T$=1 line must be equal
for any $\alpha$, implying that the peak height of $C(T)/\gamma_{\rm N}T$ is lowered when its position becomes lower $T$
to satisfy the entropy balance.
Note that no analogous constraint exists in the $T_1$ case.
Thus the peak height in $1/T_1$ is rather ``arbitrary'' in the sense that
it depends sensitively on dirtiness, impurities or lattice imperfections, etc, where
the smearing parameter $\eta$ becomes larger, making the peak height in $1/T_1$
lower.

\begin{figure}
\begin{center}
\includegraphics[width=12cm]{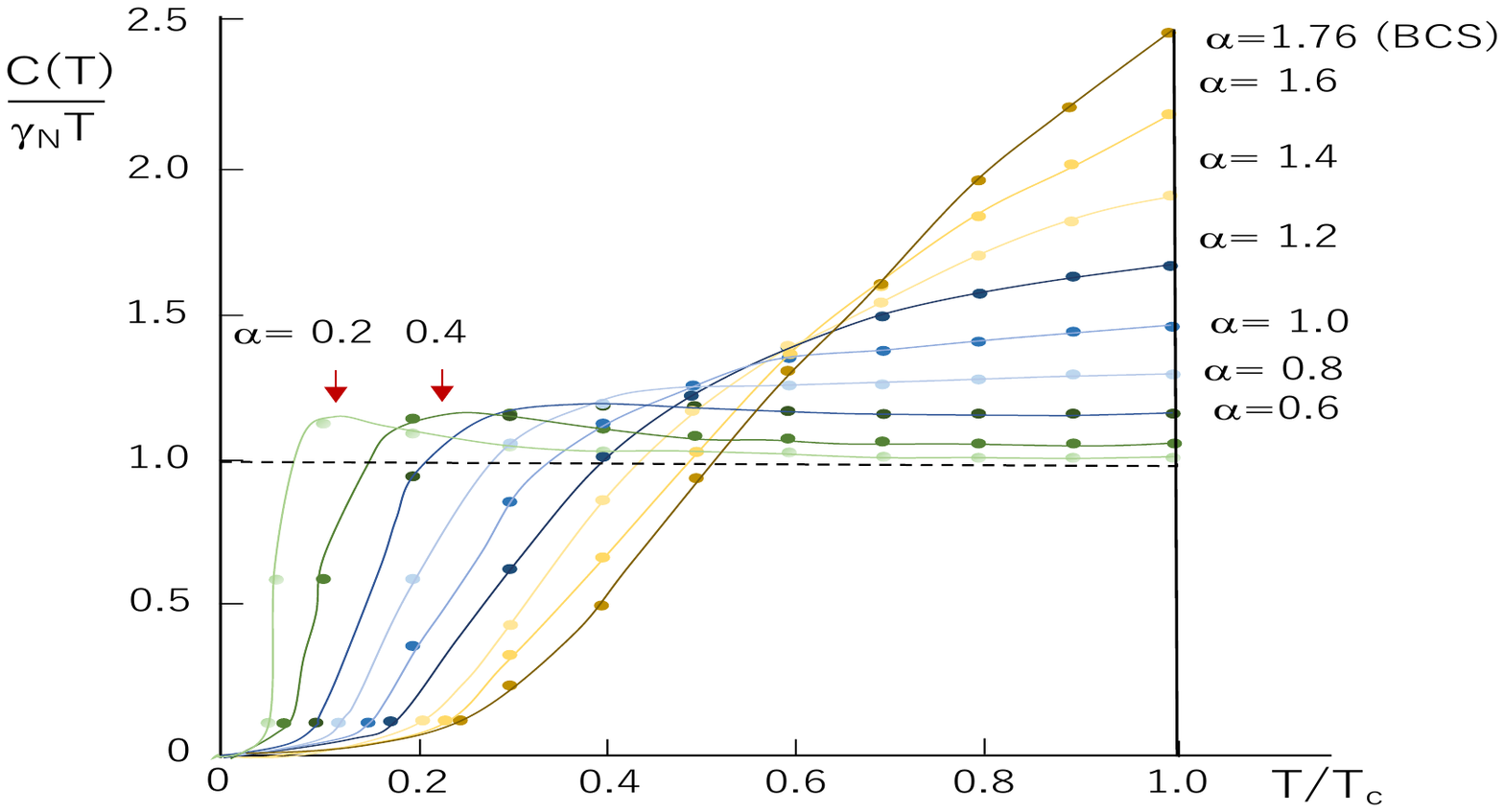}
\end{center}
\caption{Specific heat $C(T)/ \gamma_{\rm N}T$ for various $\alpha$, including the
BCS case. The jump is given by Eq.~(\ref{jump}). The maximum of $C(T)/ \gamma_{\rm N}T$
shifts well below  $T_{\rm c}$ for smaller $\alpha$ cases. The down arrows indicate the peak positions for
$\alpha$=0.2 and 0.4. The two areas above and below $C(T)/ \gamma_{\rm N}T$=1
are equal due to the entropy balance.
\label{ct}}
\end{figure}

\subsection{Generic properties of condensates with an extremely small superfluid density}

It is evident from the above analyses on $T_1$ and $C(T)$ that
the second order phase transition associated with condensates having the extremely small 
superfluid density characterized by the smaller 
$\alpha=\Delta_0/k_{\rm B}T_{\rm c}$ is harder to detect 
because the specific heat jump corresponding to the derivatives of the entropy 
with respect to temperature or magnetic field becomes small and the peak height in $1/T_1$ 
is shifted to lower temperatures and becomes much less pronounced.
As we show in the following sections that these features indeed are shared with and found in UTe$_2$ 
as the ``hidden'' second phase.

We also point out the close analogy
 between the small $\rho_{\rm s}$ condensate system and the minor band condensate
 in multiband superconductors. The latter is characterized by low carrier density, or small DOS,
thus it is a small $\rho_{\rm s}$ condensate system.
These systems share common physical properties and there is no essential distinction  between them.
UTe$_2$ is characterized by having two bands consisting of $\alpha$-FS and $\beta$-FS, making 
the data analysis more involved but also reveals richer physics.

\section{Hidden second transition at low fields and temperatures}

Having established the theoretical framework, we now analyze the experimental data to demonstrate the existence of a hidden second superconducting phase around $T_{\rm c2}$$\sim$0.3K under ambient pressure.
This finding leads us to provide a new perspective on the whole phase
diagrams over the $H$-axis and $P$-axis. 

The hidden superconducting phase $A_2$, embedded within the high-temperature $A_1$ phase,
provides the key to understanding various seemingly unrelated experimental puzzles
 in UTe$_2$. For example, as mentioned in the Introduction,
the high field phase SC2 characterized by the low phase stiffness is now identified as the same phase $A_2$
and can be understood as the field-induced re-emergence
 of the  $A_2$ phase driven by an applied field $H$$\parallel$$b$-axis.
Furthermore, this phase diagram with the reentrant high field phase $A_2$ is predicted to exist also for $H$$\parallel$$a$-axis
and for $H$$\parallel$$c$-axis.
The naming of $A_1$ and $A_2$ is used to distinguish the two phases,
where the detailed characterizations are explained later.

In order to uncover the origin of the mysterious peaks in $1/T_1T$ around $T$=0.1K for $H$$\parallel$$b$-axis
at $H$=0.38T, 0.65T, and 0.80T~\cite{matsumura0}, we first notice the following experimental facts: \\
\noindent
(1) Just below $T_{\rm c}$=2.1K, $1/T_1T$ shows a strong decrease without any signature of 
the Hebel-Slichter enhancement, indicating that the high $T$ phase is described by having
a conventional value of $\alpha$
 as observed in many unconventional SC~\cite{cecu2si2,maclaughlin,kitaoka}.\\
\noindent
(2) This strong decrease of $1/T_1T$ stops at around 0.3K because
$1/T_1T$ reaches the ``residual'' DOS regime as seen from Fig.~\ref{line2}. 
The estimated ``residual'' DOS$\sim$12mJ/mol$\cdot$K$^2$
whereas the normal DOS $\sim$120mJ/mol$\cdot$K$^2$.\\
\noindent
(3) Thus it is expected that $1/T_1T$ would be expected to exhibit a $T$-independent 
constant corresponding to the square of this
``residual'' DOS value below 0.3K in UTe$_2$ (see Fig.~\ref{line2}).\\
\noindent
(4) Experimentally, however, $1/T_1T$ gradually rises, develops a broad peak and decreases again upon further cooling.
The peaking $T$ shifts to a higher $T$ and the height becomes lower as $H$ increases.
The peak height corresponds to DOS$\sim$28mJ/mol$\cdot$K$^2$.

These observations can be naturally explained by assuming the existence of 
a second superconducting phase, $A_2$,
whose transition temperature is at $T_{\rm c2}$$\sim$0.3K whereas the $A_1$ phase is responsible for the high $T$ phase
at $T_{\rm c}$=2.1K.

In the high $T$ region the $A_1$  phase exists while below $T_{\rm c2}$ the
 $A_1$  and $A_2$ phases coexist.
Accordingly, the previous analysis must be extended to the coexistence of two condensates. 
In this case $1/T_1T$  is given by the formula~\cite{ishida1}:

\begin{eqnarray}
{(T_1T)_{\rm N}\over (T_1T)}=n^2_{A_1}\Big({(T_1T)_{\rm N}\over (T_1T)}\Big)_{A_1}+n^2_{A_2}\Big({(T_1T)_{\rm N}\over (T_1T)}\Big)_{A_2}
\label{t11}
\end{eqnarray}

\noindent
where $n_{A_1}$($n_{A_2}$) denotes the fractional DOS associated with
 the $A_1$ ($A_2$) condensate with $n_{A_1}+n_{A_2}=1$.
 The quadratic weights arise because the relaxation rate is proportional to the square of the quasiparticle density of states.
The total relaxation rate is therefore obtained as the weighted sum of the contributions from the two condensates.

\subsection{$H$$\parallel$$b$-axis}

\begin{figure}
\begin{center}
\includegraphics[width=12cm]{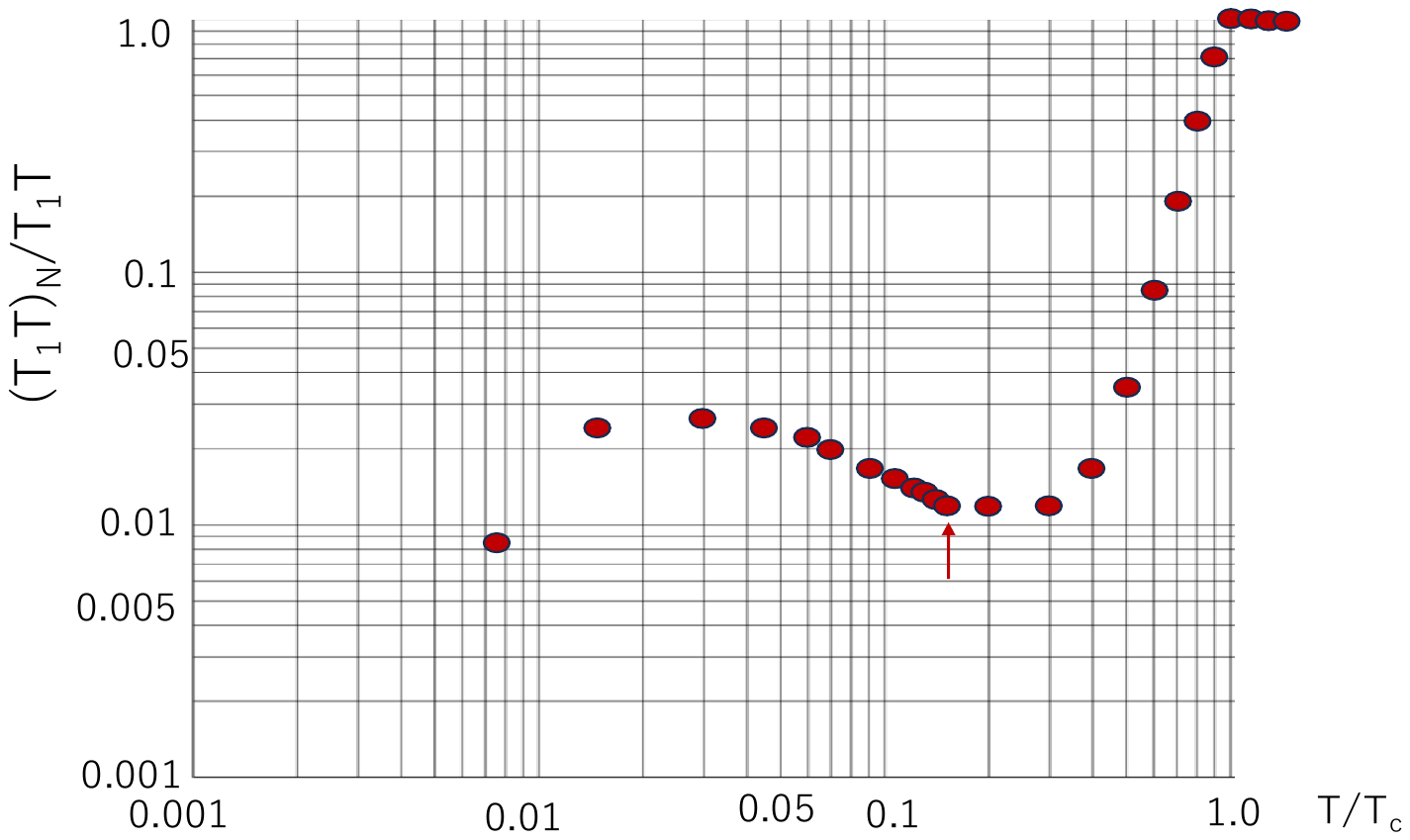}
\end{center}
\caption{$(T_1T)_{\rm N}/T_1T$ for the two fully gapped condensates  in low magnetic fields,
showing a characteristic low-temperature peak. The upward arrow indicates
the second transition temperature at $T_{\rm c2}$=0.3K.
Parameters used are $n_{A_1}$=0.9, $n_{A_2}$=0.1, $\alpha_{A_1}$=3.5,
$\alpha_{A_2}$=0.2, $\eta_{A_1}=5\times10^{-2}$, 
and $\eta_{A_2}=5\times10^{-5}$, and set $T_{\rm c2}/T_{\rm c1}$=0.15.
\label{t1}}
\end{figure}

As shown in Fig.~\ref{t1}, we calculate $1/T_1T$ normalized by its normal value
under the assumption that  both phases are fully gapped for low $H$.
We set $T_{\rm c2}$=0.3K. For $T_{\rm c2}<T<T_{\rm c}$,
$1/T_1T$ decreases rapidly  because we assume a larger $\alpha$=3.5
and eventually reaches a small value described by the exponentially activated low-temperature regime.
When the $A_2$ phase emerges at $T_{\rm c2}$, $1/T_1T$ starts to increase and 
after taking a local maximum
decreases again in further lower $T$ region.
We assume a much smaller $\alpha$=0.2 for the $A_2$ phase shown in Fig.~\ref{t1eta}
where we have chosen a small $\eta$=5$\times 10^{-5}$ to reproduce a relatively large enhancement~\cite{matsumura0}.
The overall features above are similar to those observed by Matsumura, {\it et al.}~\cite{matsumura0}
for the $H$$\parallel$$b$-axis at $H$=0.38T, 0.65T, and 0.80T.
The mysterious peak in  $1/T_1T$ for low $H$ unidentified so far can be naturally interpreted as
 the second phase transition
into the $A_2$ phase.

\subsection{$H$$\parallel$$c$-axis}

To further support this interpretation, we analyze the NMR data 
for higher magnetic fields applied along the $c$-axis.
According to the data~\cite{matsumura0}, for $H$=0.75T and 2.8T,
the rapid decrease of $1/T_1T$ below $T_{\rm c}$ levels off, 
forming a plateau, and then decreases again upon further cooling 
without exhibiting a pronounced peak.
These features are understood basically in the same manner as above.

\begin{figure}
\begin{center}
\includegraphics[width=12cm]{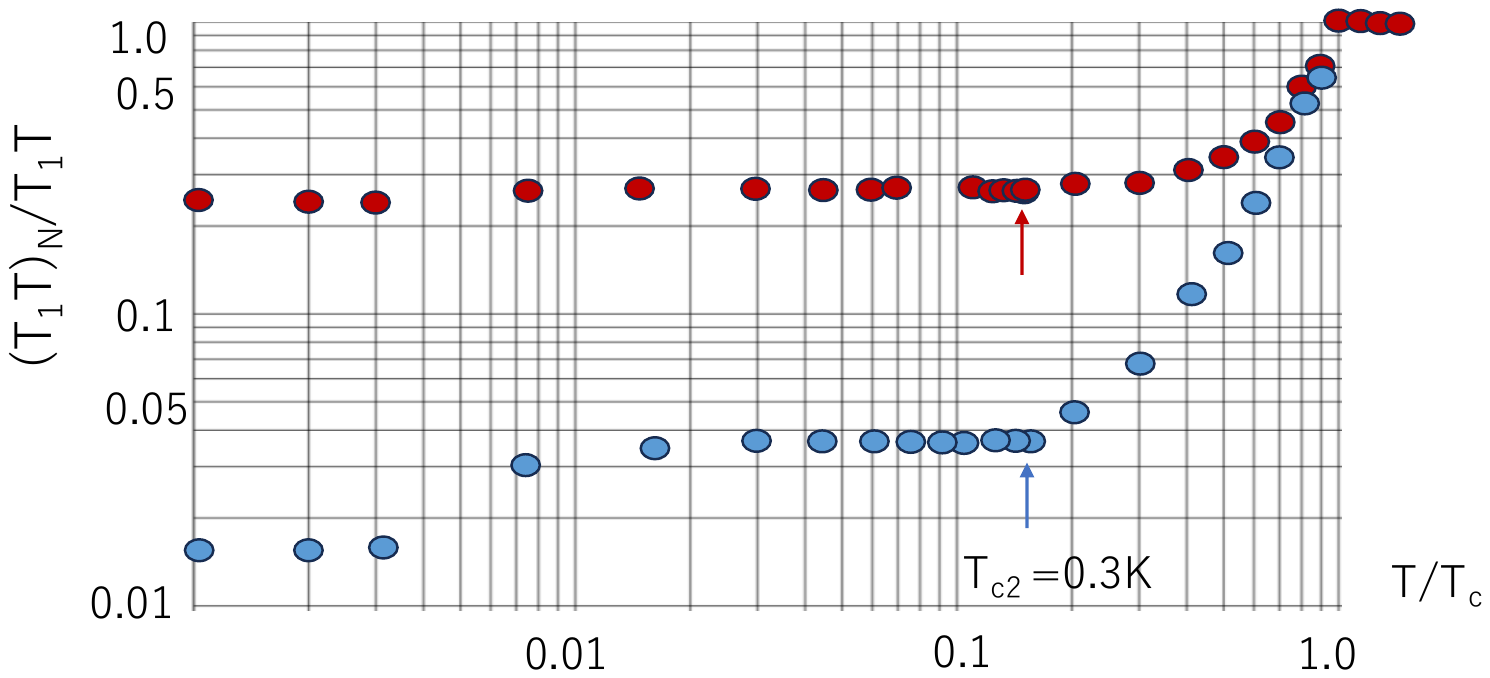}
\end{center}
\caption{$(T_1T)_{\rm N}/T_1T$ for finite magnetic fields, showing
the double shoulder structure. 
The upper (lower) curve corresponds to the higher (lower) magnetic field.
The up arrows indicate
the second transition temperature at $T_{\rm c2}$=0.3K.
The calculations use $\gamma(H)$=0.5(0.3) for upper(lower) curve
and set  $T_{\rm c2}/T_{\rm c1}$=0.15.
\label{t1c}}
\end{figure}

The difference is that the plateau value in the intermediate-temperature 
region is higher, making the contribution from the $A_2$ 
phase difficult to distinguish from that of the $A_1$ phase.
Namely the peak is is masked by the residual DOS
 of the $A_1$ phase, which corresponds to 0.16 and 0.3$\gamma_{\rm N}$
for $H$=0.75T and 2.8T, respectively.
The double-shoulder $T$ dependence of $1/T_1T$  can therefore be understood 
as due to the double transition from $A_1$ to $A_2$, indicating that 
the $A_2$ phase exists at $T_{\rm c2}$=0.3K.
We illustrate it in Fig.~\ref{t1c} by calculating $1/T_1T$ for finite $H$ cases.
The double shoulder structure in $1/T_1T$ is clearly seen for the lower-field case (lower curve)
while it is buried in the upper curve for the higher field case.

\begin{figure}
\begin{center}
\includegraphics[width=12cm]{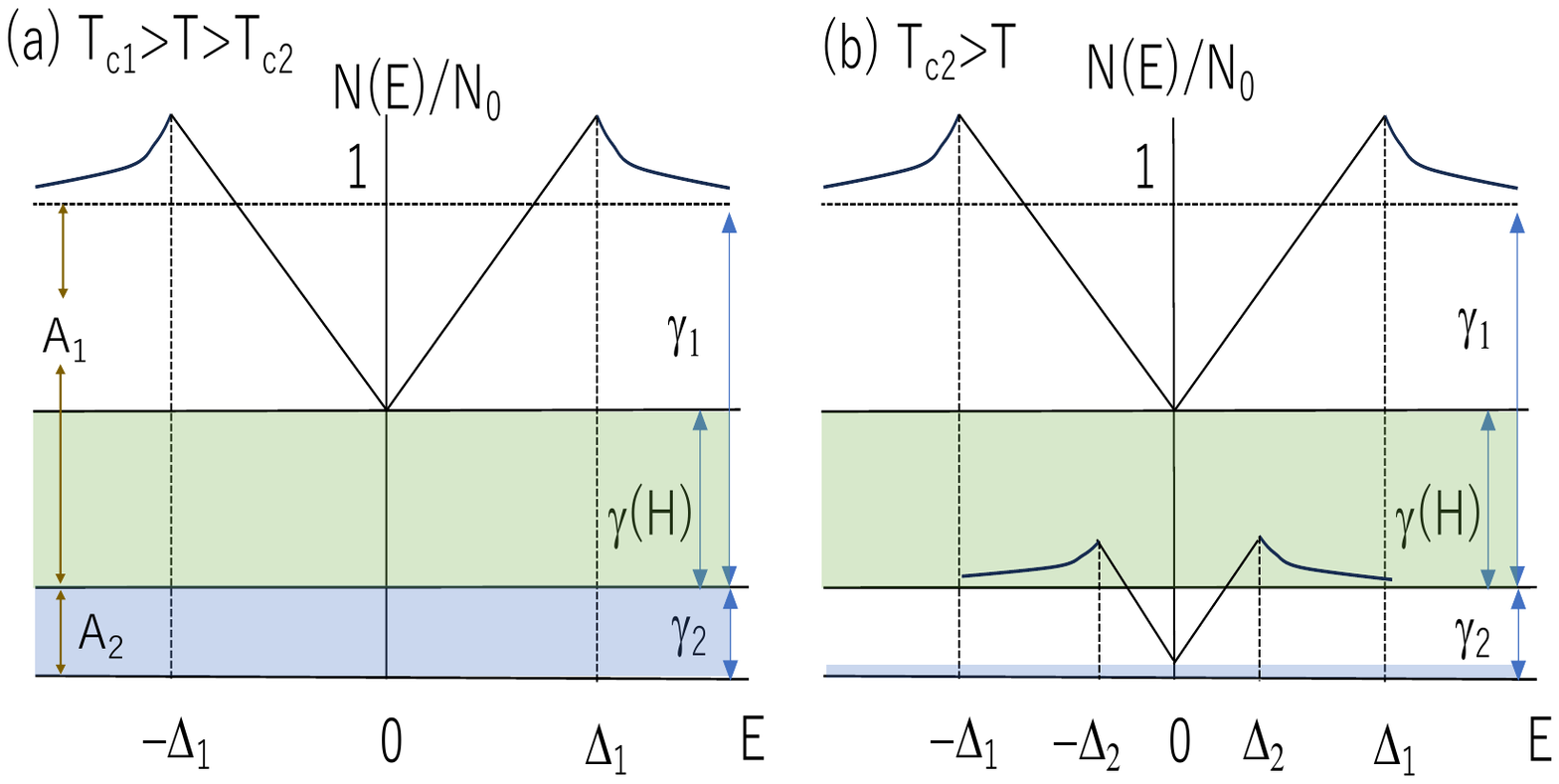}
\end{center}
\caption{
Schematic density of states in (a) the intermediate-temperature region, 
$T_{\rm c1}$$>$$T$$>$$T_{\rm c2}$, and (b) the low-temperature region, $T$$<$$T_{\rm c2}$.
The total DOS is $\gamma_1$ for the $A_1$ plus $\gamma_2$ for the $A_2$.
\label{2banddos}}
\end{figure}

The double shoulder structure in $1/T_1T$ is physically understood as follows:
The DOS in the temperature range: $T_{\rm c2}$$<$$T$$<$$T_{\rm c1}$ is shown schematically in 
Fig.~\ref{2banddos}(a). The DOS for the $A_1$ whose total normal-state DOS is denoted by
 $\gamma_1$ is gapped and exhibits a V-shaped DOS with a finite zero-energy value  $\gamma(H)$
 whereas the $A_2$ band, with DOS $\gamma_2$, remains in the normal-state.
Therefore, starting from $T_{\rm c1}$ $1/T_1T$ drops sharply until reaching the bottom $\gamma(H)$.
In the intermediate $T$ region $1/T_1T$ shows a plateau.
Then further below $T_{\rm c2}$ shown in Fig.~\ref{2banddos}(b) the gap for the $A_2$ opens,
giving rise to the second shoulder structure in $1/T_1T$.

\subsection{Specific heat signature for the $A_2$ phase at H=0}

According to the experiments by Totsuka, {\it et al.}~\cite{totsuka},
the specific heat at $H$=0 exhibits a shallow minimum at around $T$$\sim$0.3K
before the pronounced upturn arising from the nuclear Schottky contribution or some unknown origin.
At the minimum $C(T)/\gamma_{\rm N}$$\sim$0.05, which is still above the residual
DOS value. Thus it could be a signature of the second transition to the $A_2$ phase.
It is not unreasonable since the specific heat anomaly of this transition to the extremely small DOS phase is
characterized by the small specific heat jump $\Delta C(T_{\rm c2})/\gamma_{\rm N}T_{\rm c2}$=0.018
estimated by Eq.~(\ref{jump}) with $\alpha_{A_2}$=0.2 and the overall $T$ dependence of $C(T)/\gamma_{\rm N}$ 
is plotted in Fig.~\ref{ct}.

\section{Construction of overall phase diagrams under $P$=0}

Having established the existence of the second superconducting phase below 
$T_{\rm c2}$$\sim$0.3K at zero field, we now construct the overall phase diagrams for ambient pressure.
This hidden $A_2$ phase survives up to magnetic fields of a few tesla, depending on the field orientation.
This naturally suggests that 
the $A_2$ reappears at finite $H$ because this phase is energetically favored under a finite $H$
due to its Cooper-pair spin aligned parallel to the applied field. Note that
 the pair spin of the $A_1$ is antiparallel to it.
Here we explore the whole phase diagrams for $H$$\parallel$$a$-axis, $b$-axis,
and $c$-axis by combining the available experimental data with the present theoretical framework.

Our consideration is based on the non-unitary spin-triplet pairing, which is governed by the
underlying magnetization $M_i$ for the $i$-direction through the magnetic coupling term given by
$\kappa_i M_i(H_i)S_i$ with the $S_i$ Cooper pair spin moment for the $i$-direction~\cite{machida6,machida7}.
Here the magnetization $M_i$ mainly comes from the localized moment of U atoms.
We have extended our theory~\cite{machida4,machida5,machida6,machida7}
 to include anisotropic coupling constants  $\kappa_i$.
The resulting phase diagrams depend sensitively on the field orientation.
We construct the phase diagram using the equation to determine $H_{\rm c2}$:

\begin{eqnarray}
{H_{{\rm c2},i}^{(\pm)}(T)=\alpha_{i}(T_{\rm c0}\pm \kappa_i M_i(H_{{\rm c2,}i}^{(\pm)}(T))-T)
}
\label{hc22}
\end{eqnarray}

\noindent
with $\alpha_{i}$ the initial slope of $H_{{\rm c2},i}$ for the $i$-direction
and $T_{\rm c0}$ transition 
temperature for $H$=0.
This equation implies that when the Cooper spin points to the field direction,  
$H_{{\rm c2},i}^{(+)}$ is enhanced while $H_{{\rm c2},i}^{(-)}$ is suppressed when it is 
antiparallel to the field direction. 
Consequently, the $A_2$ phase may reappear
at a higher field. In the following we will see that in a higher
field the $A_2$ is bound to reappear for all three directions
because the spin-orbit coupling that pins the Cooper-pair spin to the crystal lattice is relatively weak. The external magnetic field can therefore gradually depin the spin polarization and align it with the applied field.

Equation~(\ref{hc22}) can be rewritten at $T$=0 as

\begin{eqnarray}
|{H_{{\rm c2},i}^{(\pm)}\mp\alpha_i \kappa_iM_i(H_{{\rm c2,}i}^{(\pm)})|=\alpha_iT_{\rm c0}.
}
\label{hc2}
\end{eqnarray}

\noindent
We define the effective field by the left hand side 

\begin{eqnarray}
{H^{(\pm)}_{{\rm eff},i}=H_{{\rm c2},i}^{(\pm)}\mp\alpha_i\kappa_iM_i(H_{{\rm c2},i}^{(\pm)})
}
\label{heff}
\end{eqnarray}

\noindent
must remain below $\alpha_i T_{\rm c0}$ for a finite solution of  $H_{{\rm c2},i}^{(\pm)}$
to exist.

\begin{figure}
\begin{center}
\includegraphics[width=12cm]{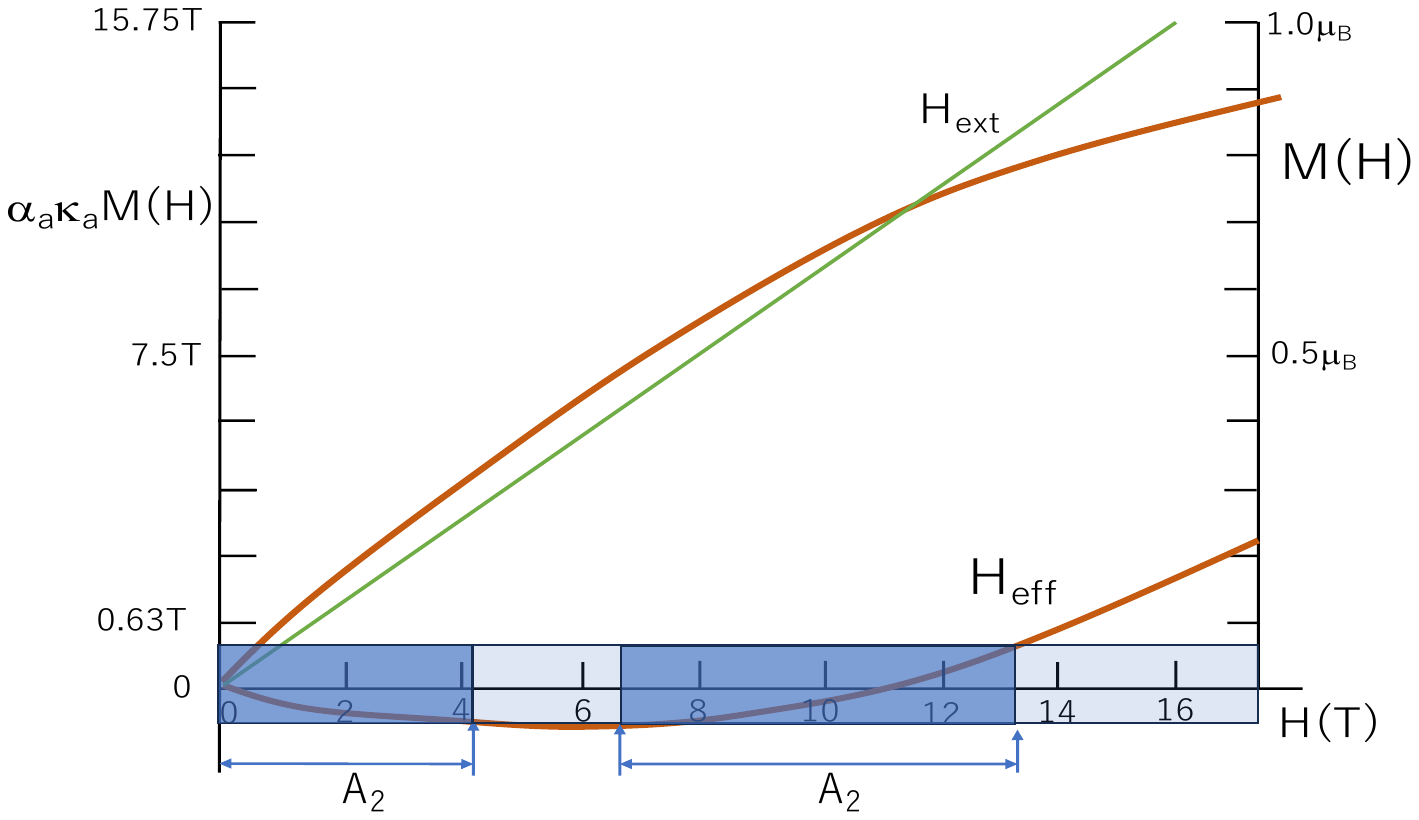}
\end{center}
\caption{Construction of the reentrant phase diagram for $H$$\parallel$$a$-axis at $T$=0.
The magnetization $M(H)$ (right-hand axis in the unit of $\mu_{\rm B}/U$ for this direction and
$\alpha_a\kappa_aM(H)$ (left-hand axis in the unit of T) are plotted.
$H_{\rm eff}=H_{\rm ext}-\alpha_a\kappa_aM(H)$ is shown as a function of $H$
together with the  allowed region defined by $\pm\alpha_aT_{\rm c2}$.
\label{hefffig}}
\end{figure}

\subsubsection{$H$$\parallel$$a$-axis}

We start with the magnetic easy $a$-axis at $T$=0.
As shown Fig.~\ref{hefffig} where $M_a(H)$~\cite{miyake} multiplied by $\alpha_{a}\kappa_a$
and $H_{\rm ext}$ are plotted, $H^{(+)}_{{\rm eff},a}$ lies within the allowed region
for  0$<$$H_{\rm ext}$$<$4T and 5T$<$$H_{\rm ext}$$<$12.5T, defining the stability region of 
 the $A_2$ phase.
$\alpha_a$=7(T/K) and $\kappa_a$=2.25(T/$\mu_{\rm B}$) 
are adopted and  $T_{\rm c2}$=0.1K is chosen to reproduce the observed value of $H_{\rm c2}$$\sim$12T.
Thus at $T$=0 the $A_2$ phase terminates once at $H$=4T and reappears at $H$=5T,
persisting up to $H$=12.5T where magnetization $M_a$ tends to saturate~\cite{miyake}
and can no longer compensate for the external field anymore.

This reentrant behavior originates from the magnetization $M_a(H)$, which is strongly nonlinear 
in the low $H$ region characteristic in the magnetic easy $a$-axis~\cite{miyake}.
This nonlinear magnetization curve ultimately gives rise to the reentrance.
Moreover the spin polarization ${\bf S}$ is pinned parallel to the $a$-axis for 
the $A_2$ phase while antiparallel for the $A_1$ phase.
Therefore,  this reentrance is unrelated to the d-vector rotation,
which is the main driving factor of the reentrant behaviors in the $b$ and $c$-axis
as we will see shortly.

The $A_1$ phase with $T_c$=2.1K is characterized by the spin polarization antiparallel to
the field direction, and corresponds to $H_{\rm c2,a}^{(-)}(T)$ in Eq.~(\ref{hc22}),
where $\alpha_a(A_1)$=15(T/K) is employed as observed~\cite{aoki}, keeping the same $\kappa_a$ value as above.
Then we obtain $H_{{\rm c2},a}(A_1)$=8.4T, which is quite reduced by the negative effect 
in $H_{\rm eff}$ of Eq.~(\ref{heff}).

Since $H_{\rm c2}(T)$ is determined by Eq.~(\ref{hc22}), it is straightforward to construct the phase
diagram containing both $A_1$ phase and $A_2$ phase where $H_{\rm c2}(0)$ and the corresponding
$T_{\rm c}$ values are connected by a straight line.
In Fig.~\ref{pd-abc}(a),
we display the resulting phase diagram for $H$$\parallel$$a$-axis.

The $A_2$ phase occupies an isolated region around the origin $T$=0 and $H$=0,
and reappears above 6T, finally disappearing at 12T. The 
horizontal transition line separates the $A_1$ and the $A_2$
phases. Therefore the point at which the four second-order transition lines meet 
constitutes a tetracritical point (TCP). Note that the $H_{\rm c2}$ line starting from 12T at $T$=0
reaches $T_{\rm c2}(H_{\rm c2})$=$T_{\rm c2}+\kappa_aM_a(H_{\rm c2})=0.3K(0.1K)+2.25\times0.72\mu_{\rm B}$=1.92K(1.72K), demonstrating that the procedure for the phase diagram construction is largely self-consistent, although it is not yet quantitatively accurate.

In contrast the $A_1$ phase extends from $T_{\rm c}$=2.1K
up to 8T. Since the spin polarization ${\bf S}$ of the $A_1$ phase
is antiparallel to the $a$-axis,
or to the magnetization vector ${\bf M}_a$,
$H^{(-)}_{\rm c2}$ is suppressed by the applied magnetic field $H$.

So far the experimental data to support the present phase diagrams are as follows:

\noindent
(1) For the low field $A_2$, Totsuka, {\it et al.}~\cite{totsuka} observed the kink in the $\gamma(H)$
curve at  $H_k$$\sim$1T at $T$=0.3K. Since the kink in entropy $S(H)$$\propto$
$\gamma(H)$ implies a second order phase transition, which is naturally identified as the
phase boundary between $A_1$ and $A_2$. This point is indicated by the green triangle in Fig.~\ref{pd-abc}(a).

\noindent
(2) Tokiwa, {\it et al.}~\cite{tokiwa0} find a horizontal phase boundary via magneto-caloric effect (MCE) and the
kink $H_k$=5.5T of $\gamma(H)$ at $T$=0.4K, which is clearly different from that by Totsuka, {\it et al.}~\cite{totsuka}.
MCE exhibits a phase transition at $H_k$ where the metamagnetic transition occurs.

\noindent
(3) The magnetization measurements by Shimizu, {\it et al.}~\cite{shimizu} revealed a pronounced dip 
structure at 5.5T in 0.3K. This also supports the present scenario
because it can be regarded as a rounded kink structure, suggesting
the discontinuity of the first derivative of the magnetization, a signature of the
second order transition.

\noindent
(4) According to Lee, {\it et al.}~\cite{roman} and Lee~\cite{sangyun} who reanalyzes the original data set in
Ref.~[\onlinecite{roman}] the kinks in $\gamma(H)$ are found to be nearly $T$ independent. 
These kink points are depicted in Fig.~\ref{pd-abc}(a).

\noindent
(5) As pointed out by Tokiwa, {\it et al.}~\cite{tokiwa0}, $H_{{\rm c2},a}(T)$ has a kink, which 
may correspond to TCP in our terminology. The present theory naturally accounts for
 the origin of the kink. 

\noindent
(6) According to Shimizu, {\it et al.}~\cite{shimizu}, there reported a weak anomaly in the M(H)
curve around 8T, suggesting the fourth internal transition line terminating at TCP.
This deserves further investigation to confirm its existence.

Theoretically, the $A_1$-$A_2$ transition is a second order phase transition.
If the matamagnetic transition accidentally coincides  with the reentrant transition
it could become first order one, as suggested by Tokiwa, {\it et al.}~\cite{tokiwa0}.
However, this scenario appears unlikely because the matamagnetic anomaly in $M_a(H)$
is barely visible on the scale of Fig.~\ref{pd-abc}(a).
 
 Although the metamagnetic transition itself does not drive the $A_1$-$A_2$ transition, the subtle nonlinearity of the magnetization curve plays a crucial role in producing the reentrant behavior discussed above. In contrast, the magnetization curves for the $b$- and $c$-axes are essentially linear at low fields and therefore cannot induce reentrance by themselves. Instead, the reentrant transition for these field directions is driven by the $d$-vector rotation, which aligns the Cooper-pair spin $\bf S$ parallel to the magnetization
 as we will see shortly.
  
  \subsubsection{$H$$\parallel$$b$-axis}

The phase diagram for $H$$\parallel$$b$-axis has been discussed in detail previously~\cite{machida3,machida5,machida6}.
Here we briefly revisit it with newly optimized set of parameters to better reproduce the
experimental data and further provides further insight into their physical implications.
According to the Knight shift experiments~\cite{ishida1} which will be discussed in detail shortly, 
$\chi_{\rm s}$ decreases below $T_{\rm c}$
in  the lower field region, and gradually approaches the normal-state value around 14T, meaning that
$\bf S$ pointed antiparallel to the $a$-axis  at $H$=0  gradually rotates and becomes aligned parallel
 to the $b$-axis
around $H$=14T. There, $\bf S$ is locked to the $b$-axis parallel to the magnetization
 ${\bf M}_b$  to gain the magnetic energy through -$\kappa_bM_b\Delta^2_{\uparrow}$.

By using the newly optimized parameters
 we construct the $H$-$T$ phase diagram for $H$$\parallel$$b$-axis
shown in Fig.~\ref{pd-abc}(b):
As for the $A_2$ phase the following parameters are used
$\alpha_b(A_2)$=35(T/K)
and $\kappa_b$=2.0(K/$\mu_{\rm B}$).
Using the experimental magnetization $M_b(H)$=0.0125 $H$(T) ($\mu_{\rm B}$)~\cite{miyake},
$H_{\rm c2}(A_2)$ is evaluated as $H_{\rm c2}(A_2)$ = 
$\alpha_b(A_2)$($T_{\rm c2}$+$\kappa_bM_b(H_{\rm c2})$)
=80T, in practice, the superconducting phase is preempted by the metamagnetic transition at $H_b$=34T.
The transition line between the $A_2$ and the normal phase is 
given by $T_{\rm c2}(H)$=$T_{\rm c2}$+$\kappa_bM_b(H)$, and emerges above $H$=14T.
Consequently, $H_{\rm c2}(A_2)$ exhibits a pronounced positive slope, 
which is absent for $H$$\parallel$$a$-axis
and $c$-axis.
Note that $T_{\rm c2}(H_{\rm c2})$=$T_{\rm c2}$+$\kappa_b$$M_b(H_{\rm c2})$=2.3K,
which is consistent with the positive slope of $H_{\rm c2}$ there.

As for the $A_1$ phase  using $\alpha_b(A_1)$=20(T/K) as observed~\cite{aoki},
$M_b(H=40T)$ = 0.5($\mu_{\rm B}$), and
the reduction $\Delta H$=$\alpha_b(A_1)$$\kappa_b$$M_b(H=40T)$ =20T,
we find $H_{\rm c2}(A_1)$$\sim$20T, which  is in good agreement with
 the upper boundary of the $A_1$ phase
as shown in Fig.~\ref{pd-abc}(b).

The horizontal transition line at 14T in Fig.~\ref{pd-abc}(b)
was originally identified by the transport measurements by Sakai, {\it et al.}~\cite{sakai}.
Recently it has also been observed thermodynamically through ultrasound measurements~\cite{valiska}. 

According to Matsumura, {\it et al.}~\cite{matsumura0}, the $1/T_1$ anomalies are observed,
which we identified these as a signature of the phase transition from
the $A_1$ to $A_2$ in the previous section. These data are now added to the new points in
Fig.~\ref{pd-abc}(b).
We also added the experimental data points coming from the kink points in $\gamma(H)$ observed
by Lee, {\it et al.}~\cite{roman,sangyun}, which constitute strong support to assign these as 
the phase boundary between the $A_1$ and $A_2$ phases.
These newly added points further corroborate our identification.

It is remarkable to see the reconstructed phase diagram shown in Fig.~\ref{pd-abc}(b)
as a whole. We understand that the positively sloped $H_{\rm c2}$ comes from 
the transition between the $A_2$ and the normal-state.
Rosuel, {\it et al.}~\cite{rosuel} detected two successive specific heat jumps
above 14T, one coming from the $A_2$ and normal phase transition and 
the $A_1$ and the $A_2$ phase transition. According to their data~\cite{rosuel},
the specific heat jump at  the $A_2$ and normal phase is smaller than
that at the $A_1$-$A_2$ phase transition.
This is consistent with our identification that the $A_2$ phase
has a small superfluid density. 
However, one intriguing issue remains. Above the tetracritical point, $H_{\rm TCP}=14$ T, the four phase boundaries meeting at the tetracritical point can, in principle, be reconnected in different ways. Consequently, it would also be possible to assign the larger superfluid density to the $A_2$ phase with ${\bf S}\parallel b$, analogous to the cases of $H\parallel a$ and $H\parallel c$. The reason why this alternative scenario is apparently not realized remains an open question for future investigation.

\subsubsection{$H$$\parallel$$c$-axis}

Finally, we construct the phase diagram for $H$$\parallel$$c$-axis.
Following the same procedure as above, we adjust the parameters used,
so as to reproduce the observed $H_{\rm c2}$ and its change in slope at $H$=6T.
In particular, we employ the new value for $\kappa_c$=4.7(K/$\mu_{\rm B}$).
The KS decreases below $T_{\rm c}$ at low $H$ and becomes field independent above
6T~\cite{matsumurachi}, implying that  $\bf S$
is locked-in toward the $c$-axis direction, whose point is depicted in Fig.~\ref{pd-abc}(c). 

As for the $A_2$, we use $\alpha_c(A_2)$=9.6(T/K).
The enhancement of $\Delta H$ amounts to $\Delta H$ =$\alpha_c(A_2)$$\kappa_c$$M_c(H)$=15.3T.
Thus $H_{{\rm c2},c}(A_2)$=$\alpha_c(A_2)$$(T_{\rm c2}$+$\kappa_c$$M_c(H_{\rm c2}))$=18.2T.
As for the $A_1$ we use $\alpha_c(A_1)$=7.0(T/K) as observed~\cite{aoki}.
The reduction $\Delta H$=$\alpha_c(A_1)$$\kappa_c$$M_c(H)$=8.4T, giving
$H_{{\rm c2},c}(A_1)$=$\alpha_c(A_1)$$(T_{\rm c}$-$\kappa_c$$M_c(H_{\rm c2}))$=6.3T.

We show the resulting phase diagram
in Fig.~\ref{pd-abc}(c). We added the point at $T$=0.3K and $H$=1T, coming from the
kink point~\cite{totsuka} in $\gamma(H)$ and also another kink points at higher $H$~\cite{roman,sangyun}.
This phase diagram in Fig.~\ref{pd-abc}(c) is quite similar to that for $H$$\parallel$$a$-axis,
however, the physical reason for producing these diagrams is completely different.
The former (latter) is due to d-vector rotation (non-linear magnetization curve).
This phase diagram in Fig.~\ref{pd-abc}(c) is similar to that for $H$$\parallel$$b$-axis in Fig.~\ref{pd-abc}(b).
However, the latter has the positive sloped $H_{\rm c2}$.
The reason for this difference comes from the initial slope differences
between them, namely $\alpha_b$$>$$\alpha_c$ which makes $H_{{\rm c2},b}$ much larger than
$H_{{\rm c2},c}$, ultimately leading to the positive slope.

We remark another piece of evidence for the existence of the $T_{\rm c2}$ transition seen
from the the $H_{\rm c1}$ measurements by Ishihara, {\it et al.}~\cite{shibauchi2} where 
the anomalies in $H_{\rm c1}(T)$ can be recognized around 0.3K, in particular for 
the $H$$\parallel$$c$-axis. These anomalies observed mean that the condensation 
energy changes at 0.3K, suggesting a phase change there corresponding to the $A_1$-$A_2$
phase transition at almost zero field.

\begin{figure}
\begin{center}
\includegraphics[width=12cm]{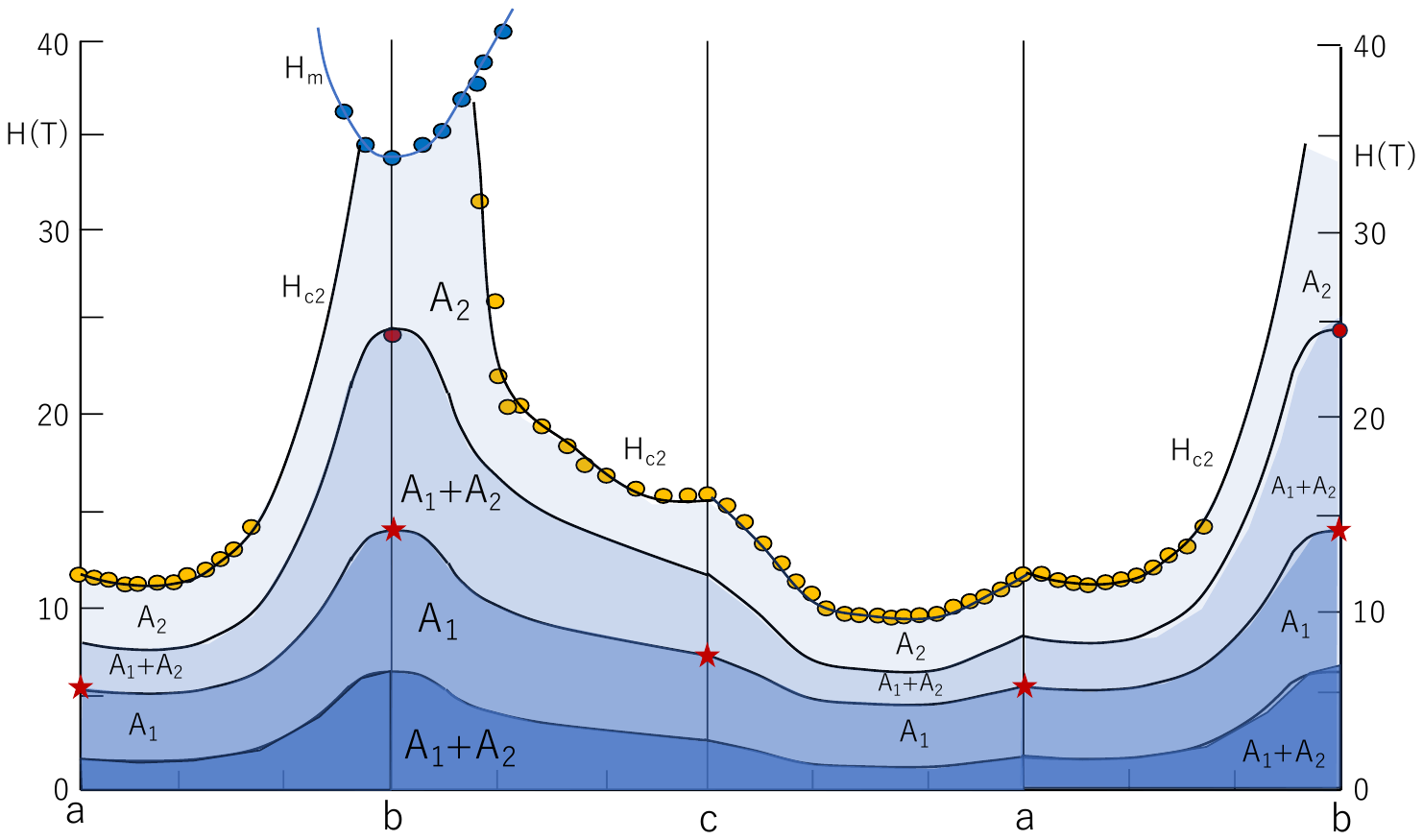}
\end{center}
\caption{Phase diagram at $T$=0 as a function of the field direction, including the three 
principal directions $a$-axis, $b$-axis, and $c$-axis.
The second order phase transitions from $A_1$+$A_2$$\rightarrow$$A_1$$\rightarrow$$A_1$+$A_2$
finally to $A_2$ with increasing field. The $H_{\rm c2}$ data come from \cite{aoki}.
$H_m$ denotes the matamagnetic transition. The star symbols show the endpoint of the
horizontal transition lines.
\label{abc}}
\end{figure}

\subsubsection{Phase diagrams for $H\parallel a$-, $b$-, and $c$-axes}

We now combine the phase diagrams for the three principal field directions to examine how the phase diagram evolves as the field direction is varied. In doing so, we ensure that the phase diagrams for the three principal axes are mutually consistent.

Figure~\ref{abc} is constructed by smoothly connecting the second-order phase transition lines at $T=0$ obtained from Figs.~\ref{pd-abc}(a), (b), and (c). Here we assume that these transition lines evolve continuously without interruption as the field direction is varied. This assumption is reasonable because the observed $H_{\rm c2}$~\cite{aoki}, also shown in Fig.~\ref{abc}, changes smoothly as a function of the field direction.

It is seen from Fig.~\ref{abc} that

\noindent
(1) The highest-field superconducting phase is always occupied by the $A_2$ phase, whose spin polarization is parallel to the applied field. This is expected because the Zeeman energy is minimized when the Cooper-pair spin $\mathbf{S}$ is aligned with the magnetic field. Consequently, the low-field $A_1$ phase, whose spin polarization points antiparallel to the $a$-axis, is eventually replaced by the $A_2$ phase as the field increases.

\noindent
(2) The horizontal transition line $H^{\star}$ in the $H$-$T$ phase diagram appears to exist for all field orientations. As shown by the red star symbols in Fig.~\ref{abc}, $H^{\star}(T=0)$ evolves smoothly from the $a$-axis to the $b$- and $c$-axes. Therefore, we expect that the $H$-$T$ phase diagram for an arbitrary field direction retains essentially the same topology, characterized by a horizontal second-order transition line.

\section{Phase diagrams under pressure}

\subsection{Ehrenfest analysis of the specific heat jumps}

Pressure experiments~\cite{pressure0,pressure1,pressure2,pressure3,pressure4,vasina,daniel,weinberger}
provide indispensable information on UTe$_2$. Here we analyze these data, focusing primarily on
specific-heat measurements in order to elucidate the nature of the $A_1$ and $A_2$ phases,
for which clear double superconducting transitions have been observed.

In particular, Vasina, {\it et al.}~\cite{vasina} measured the AC specific heat at
$P=0.06$, 0.11, 0.19, and 0.38 GPa, while Weinberger, {\it et al.}~\cite{weinberger}
performed adiabatic specific-heat measurements at
$P=0.135$, 0.379, and 0.604 GPa.

If the double transition originates from a symmetry-breaking field that splits
the original transition temperature $T_{\rm c0}$ into two successive transitions
at $T_{\rm c1}$ and $T_{\rm c2}$, an Ehrenfest relation for the corresponding
specific-heat jumps can be derived~\cite{halperin}:

\begin{eqnarray}
\frac{\Delta C_1}{\Delta C_2}
=
\frac{T_{\rm c1}}{T_{\rm c2}}
\cdot
\frac{T_{\rm c0}-T_{\rm c2}}
     {T_{\rm c1}-T_{\rm c0}},
\label{ehrenfest}
\end{eqnarray}

\noindent
where $\Delta C_1$ and $\Delta C_2$ are the specific heat jumps for $T_{\rm c1}$ and $T_{\rm c2}$, respectively.
This is purely a thermodynamic relationship, independent of microscopic details.
Our analysis goes as follows: For example,
at $P$=0.79(0.38)GPa~\cite{aoki}, the observed $\Delta C_1/\Delta C_2$=1.38(0.28),
$T_{\rm c1}$=1.25(1.61)K and $T_{\rm c2}$=2.88(2.38)K, yielding $T_{\rm c0}$=1.85(1.89)K.
The $P$=0.19GPa data set gives rise to $T_{\rm c0}$=1.85K$\sim$1.78K.
Therefore, we obtain roughly the same $T_{\rm c0}$=1.85K for three pressure data in common.
There are another two sets of the specific heat jump data~\cite{pressure0,weinberger} with different $T_{\rm c}$ values
and measurement methods, yielding slightly differnt $T_{\rm c0}$ values unsurprizingly. 
However, within the same set of data, the constant $T_{\rm c0}$ results
($T_{\rm c0}$=1.1K for~\cite{pressure0} with $T_{\rm c}$=1.6K sample and 
$T_{\rm c0}$=1.6K for~\cite{weinberger} with $T_{\rm c}$=2.1K sample), strengthening
the present finding that there exists $T_{\rm c0}$ unchanged under $P$ near the critical pressure $P_{\rm TCP}$.

\begin{figure}
\begin{center}
\includegraphics[width=12cm]{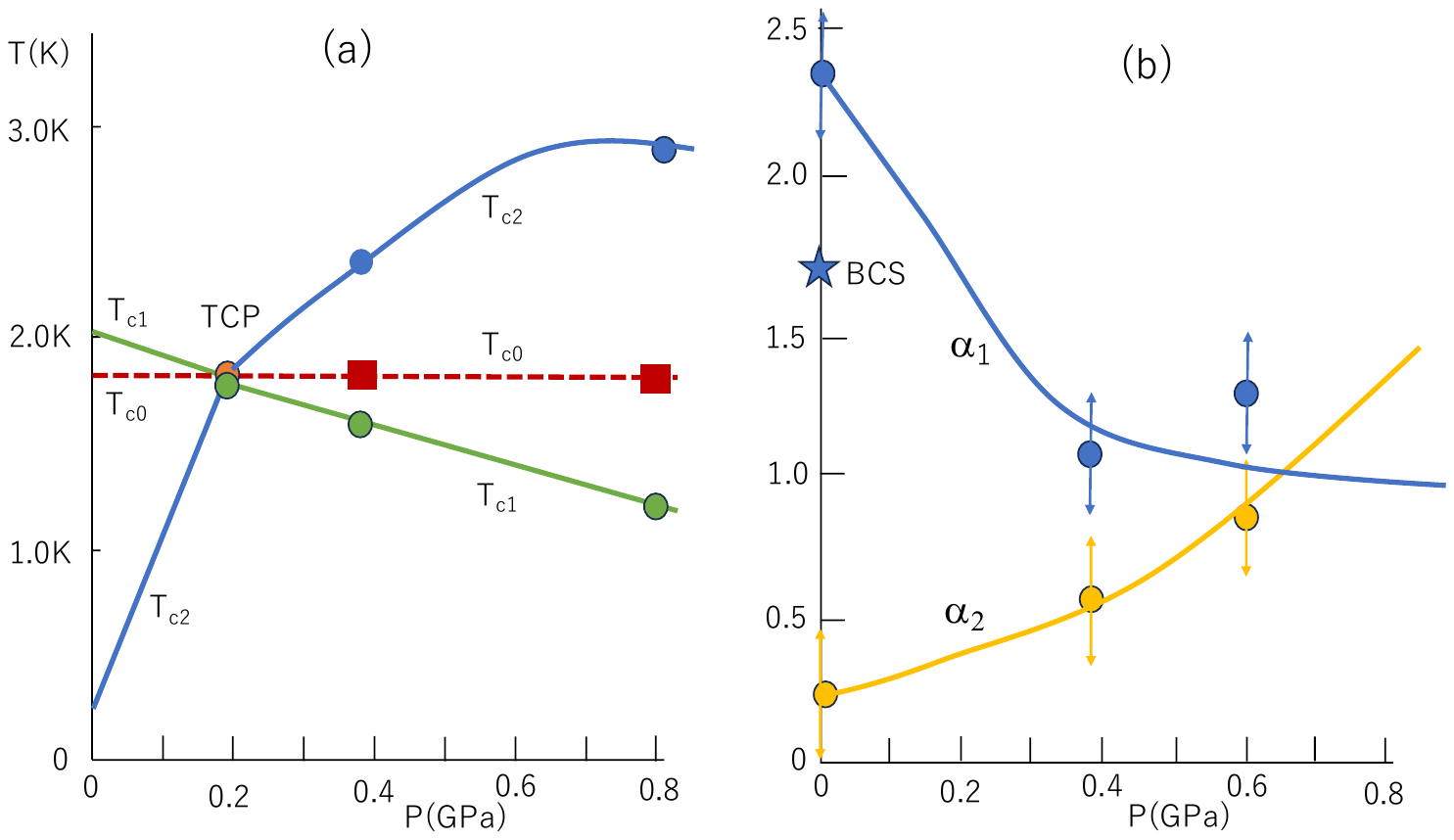}
\end{center}
\caption{(a) $T_{\rm c0}$ (squares) determined by the Ehrenfest relation Eq.~(\ref{ehrenfest})
via the specific heat jumps and the associated $T_{\rm c1}$ and $T_{\rm c2}$.
The three temperatures $T_{\rm c0}$, $T_{\rm c0}$, and $T_{\rm c0}$ meet at $P$=0.18GPa,
constituting the tetracritical point. 
(b) $P$ dependences of $\alpha_1$ and $\alpha_2$ estimated by the specific heat data~\cite{weinberger}
using Eq.~(\ref{jump}). $\alpha$=1.76 corresponds to the weak coupling BCS value.
\label{tc00}}
\end{figure}

As shown on Fig.~\ref{tc00}(a), $T_{\rm c0}$ independent of $P$ passes through the meeting point
at $P$=0.18GPa where $T_{\rm c1}$ and $T_{\rm c2}$ coincide.
Namely, it is remarkable to see that at the TCP of $P$=0.18GPa all  three observed temperatures satisfy
$T_{\rm c0}$=$T_{\rm c1}$=$T_{\rm c2}$=1.85K.

This finding leads us 

\noindent
(1) to predict the fourth second order transition line between $P$=0 and $P$=0.18GPa, in order to 
satisfy the thermodynamic stability of the phase diagram in $T$-$P$ plane.

\noindent
(2) to calculate $\Delta C_2/\Delta C_1$=0.02 in the ambient pressure, 
resulting in $\Delta C_2$=4.7mJ/molK$^2$
at $T_{\rm c2}$=0.3K by substituting the observed $\Delta C_1$=240mJ/molK$^2$~\cite{totsuka}
into Eq.~(\ref{ehrenfest}).
This small number is consistent with our claim that 
the $A_2$ phase is characterized by an extremely low superfluid density.

The data set of the specific heat experiments~\cite{weinberger} is worth analyzing further because the 
specific heats under pressure are measured by the adiabatic method, yielding
the absolute values. This allows us to determine $\alpha$ as a function of $P$
using Eq.~(\ref{jump}).
As shown in Fig.~\ref{tc00}(b), $\alpha_1$($\alpha_2$) monotonically decreases (increases).
They become equal around $P$=0.6GPa and the magnitudes are reversed beyond it.
This implies that the $A_2$ phase recovers its DOS toward the BCS value 1.76 as $P$ increases,
which is consistent with the experimental fact~\cite{matsumuragakkai} that $1/T_1$ drops below $T_{\rm c2}$ rather slowly at 
$P$=0.8GPa, implying a larger $\alpha_2$. 
We notice that $\alpha_1+\alpha_2$ is roughly a constant as a function of $P$.
This is reasonable because the total DOS for a system should be conserved.

It is quite interesting to examine the $\alpha_1$ change under $P$ by monitoring
the Hebel-Slichter peak, which is sensitive to the $\alpha$ value as discussed in Fig.~\ref{t1bcs}.
Namely, since $\alpha_1$ is relatively small just above $P_{\rm TCP}$,
the Hebel-Slichter peak coming from the full gap in the $\alpha$-Fermi surface
might be observed. The existing $1/T_1$ data at $P$=0.8GPa show a mild decrease
just below $T_{\rm c2}$=3K~\cite{matsumuragakkai}, implying that $\alpha_2$ is already
recovered at the pressure as seen from Fig.~\ref{tc00}(b).

The same analysis by the Ehrenfest relation Eq.~(\ref{ehrenfest}) is applied to the data set for the successive transition
under applied $H$$\parallel$$b$-axis case in the ambient pressure~\cite{rosuel} above $H$$>$14T,
yielding no unique value of  $T_{\rm c0}$ because $T_{\rm c0}$ is $H$-dependent and does not pass 
through the TCP shown in Fig.~\ref{pd-abc}(b).
This can be understood in terms of the different situations between them:
$P$ does not strongly perturb the underlying electronic system
while $H$ drives the system strongly as evidenced by the fact
that the effective masses change as $H$ increases~\cite{tokunaga-prl}. 
 
\subsection{Construction of the phase diagram under $P$}

Since we have identified the $A_2$ phase and located it below
$T_{\rm c2}$=0.3K at ambient pressure,
it is now straightforward to construct the phase diagram under $P$.
At $P$=0.18GPa, the four second-order transition lines meet at a tetracritical point (TCP).
The hypothetical transition temperature $T_{\rm c0}$=1.85K extends
horizontally through the TCP.
Therefore, the ground state at $T$=0 is always a mixture of the $A_1$ and $A_2$ phases,
whose relative weights vary with $P$, except at the TCP, where
$T_{\rm c0}$=$T_{\rm c1}$=$T_{\rm c2}$=1.85K and the hypothetical
$T_{\rm c0}$ becomes a real transition temperature.

At the TCP, the SO(3)$^{\rm spin}$ symmetry is nearly restored.
The genuine A phase (Anderson-Brinkman-Morel phase) is thus almost realized
in spin space.
However, the orbital part, namely the gap structure, is different.

\begin{figure}
\begin{center}
\includegraphics[width=12cm]{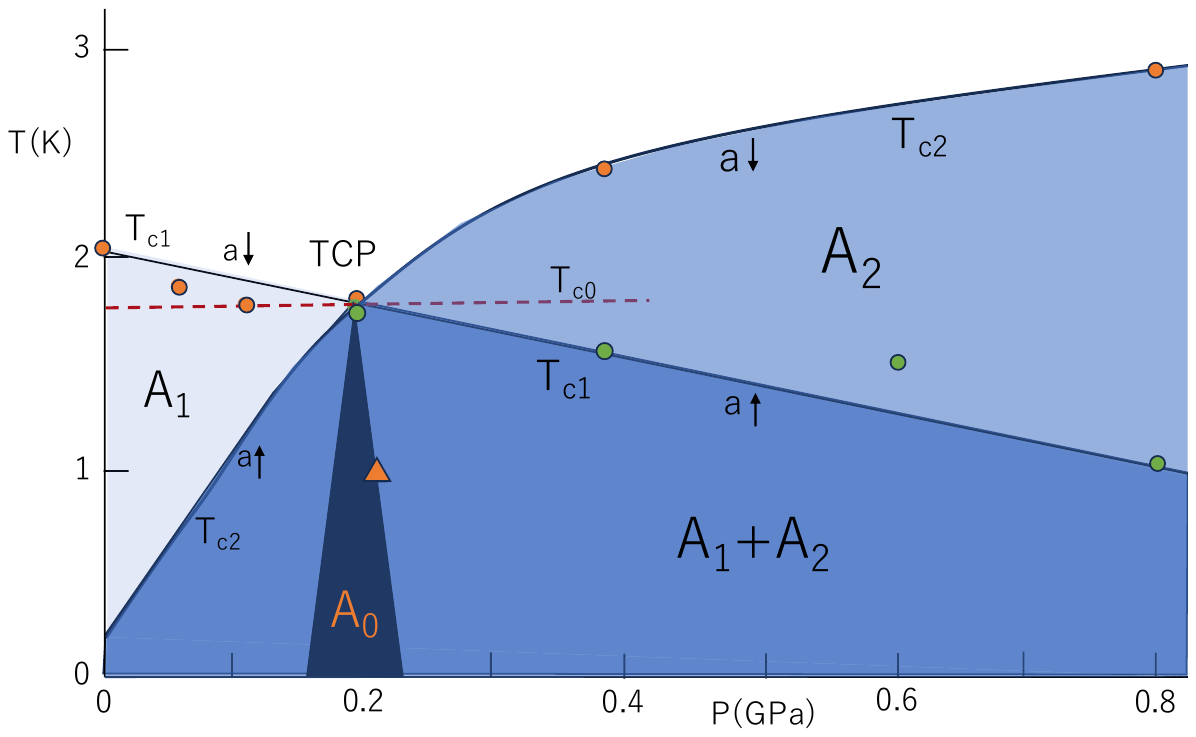}
\end{center}
\caption{Phase diagram in the $P$--$T$ plane.
The two second-order phase transition lines intersect at
$P$=0.18GPa and $T$=1.85K.
Two additional phase boundaries corresponding to the $A_0$ phase
emanate from the intersection point.
\label{pd-p}}
\end{figure}

Kamat, {\it et al.}~\cite{kamat2} recently found an additional unidentified transition
at $P$=0.21GPa and $T$=1K, marked by the triangle symbol in Fig.~\ref{pd-p},
using ultrasound measurements, and claimed it to be the missing fourth internal phase transition.
We are inclined to interpret it as a signature of the missing $A_0$ phase,
whose existence is predicted in Refs.~\cite{machida1,machida2}.
The $A_0$ phase is required to complete the full SO(3)$^{\rm spin}$ symmetry around the TCP.

As shown in Fig.~\ref{pd-p}, there are three phases, $A_1$, $A_2$, and $A_0$, in the $P$--$T$ plane.
In the highest-temperature superconducting phase below the normal-state,
the Cooper-pair spin polarization is always along the negative $a$-axis direction.

Kinjo, {\it et al.}~\cite{kinjo} reported that under $P$=1.2GPa the KS remains unchanged at
$T_{\rm c2}$ and starts to decrease below $T_{\rm c1}$ for $H$$\parallel$$b$-axis
at $H$=0.8, 1.0, and 2.5T.
This behavior is interpreted as evidence for a d-vector rotation occurring at lower fields
to gain magnetic energy.
At $H$=0, the $\mathbf{S}$ vector, which points antiparallel to the $a$-axis in the $A_2$ phase,
rotates toward the positive $b$-axis direction at $T_{\rm c2}$.
Consequently, in the $A_1$ phase, $\mathbf{S}$ becomes antiparallel to the $b$-axis,
causing the KS to decrease below $T_{\rm c1}$.

If the $A_0$ phase indeed exists and is embedded deep inside the phase diagram,
the meeting point is no longer, strictly speaking, a tetracritical point.
At present, we do not know the exact critical behavior associated with such a multicritical point,
which deserves further theoretical investigation~\cite{mermin}.

\subsection{Phase diagram for $H$$\parallel$$a$-axis under pressure}

Since the phase diagrams for the three principal field directions evolve smoothly with increasing pressure,
it is straightforward to understand their evolution in the $T$--$H$ plane.
Here we revisit our construction of the phase diagrams under pressure
(see Figs.~6, 7, and 8 of Ref.~\cite{machida7} for details).

Among them, the phase diagram for $H$$\parallel$$a$-axis deserves special attention because
the $a$-axis is the magnetic easy axis, and its spin fluctuations are expected to break the
SO(3)$^{\rm spin}$ symmetry and determine the direction of $\mathbf{S}$ to be antiparallel to the
$a$-axis at $H$=0.
Indeed, we hypothesized this scenario and confirmed it as follows.

\begin{figure}
\begin{center}
\includegraphics[width=16cm]{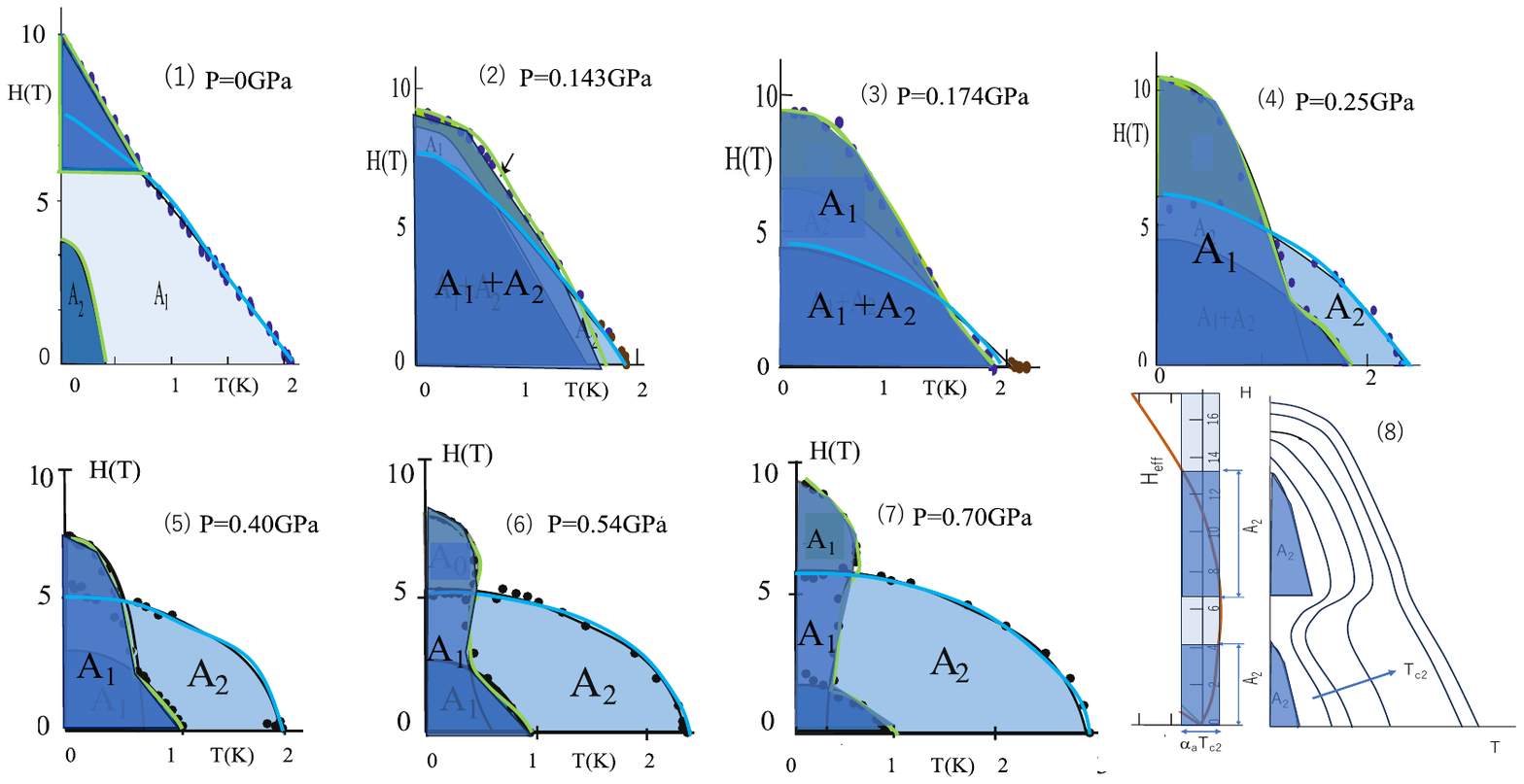}
\end{center}
\caption{Evolutions of the phase diagrams under $P$ for $H$$\parallel$$a$-axis
from (1) $P$=0 to (7) $P$=0.70GPa through the critical pressure $P$=0.18GPa. The data points
come from the references cited in \cite{machida7}.
(8) Schematic evolution of the phase diagram starting from (1) $P$=0 and upon decreasing
$P$ from (7) toward $P$=0.18GPa, the detached $A_2$ (renamed $A_1$ for $P$$>$$P$=0.18GPa) phase
emerges into the continued single phase because the allowed region defined by $\alpha_a T_{\rm c2}$
is widened as shown in left hand side in (8).
\label{ha-p00}}
\end{figure}

We predict the pressure evolution of the phase diagram for $H$$\parallel$$a$-axis
and interpret the existing experimental data shown in Fig.~\ref{ha-p00}.
As depicted in Fig.~\ref{ha-p00}, at ambient pressure the $A_2$ phase is separated into
low- and high-field regions.
As the pressure increases, $T_{\rm c2}$ increases, thereby widening the stable region of the
$A_2$ phase around the TCP.
Eventually, the two separated regions merge, and the corresponding
$H_{\rm c2}$ becomes larger than the original $H_{\rm c2}$ determined by the $A_1$ phase.

The overall phase diagrams under pressure shown in Fig.~\ref{ha-p00}
consist of an ``S''-shaped $A_2$ phase and a strongly suppressed $A_1$ phase.
The latter originates from the fact that the $A_1$ phase is characterized by
$\mathbf{S}$ pointing antiparallel to the $a$-axis, whose orientation cannot
flip toward the positive $a$-axis direction.

\subsection{GL analysis}

According to the GL theory~\cite{machida1,machida2,machida3,machida5,machida6,machida7},
the ratio of the specific-heat jumps is given by

\begin{eqnarray}
{{\Delta C_1\over \Delta C_2}={T_{\rm c1}\over T_{\rm c2}}\cdot{\beta_1\over{\beta_1+\beta_2}},
}
\label{jumphi}
\end{eqnarray}

\begin{align}
T_{\rm c1} &=T_{\rm c0}+\kappa M^{(0)}_a \nonumber \\
T_{\rm c2}  &=T_{\rm c0}-\kappa M^{(0)}_a{{\beta_1-\beta_2}\over {2\beta_2}}.
\label{tc12}
\end{align}

\noindent
From Eq.~(\ref{tc12}), we obtain
$\beta_1/\beta_2=13.5$ and
$\kappa M^{(0)}_a=0.24$ K
for $T_{\rm c1}=2.1$ K,
$T_{\rm c0}=1.85$ K,
and $T_{\rm c2}=0.3$ K.
These values yield
$\Delta C_2/\Delta C_1\sim0.14$,
which is somewhat larger than our previous estimate of
$\sim0.02$, but is nevertheless consistent with the expected trend.
By contrast, the weak-coupling result for a spherical Fermi surface gives
$\beta_1/\beta_2=-2.0$,
indicating that the present system is in the strong-coupling regime.

\section{Knight shift: How to identify the spin rotation}

It is true that the actual phase transition associated with the Knight shift (KS)
occurs through a change in the vortex configuration, where the spin texture
${\bf S}({\bf r})$ assumes different spatial patterns~\cite{tsutsumi}.
Therefore, the spatial average $\langle{\bf S}({\bf r})\rangle$ is the quantity
probed in KS measurements.

The KS provides information on the spin susceptibility $\chi_s$, which is the key quantity
for identifying the underlying spin texture and, ultimately, the pairing symmetry of a
spin-triplet superconductor.
In the spin-singlet case, $\chi_s$ arises exclusively from quasiparticles thermally excited
or field-induced out of the condensate, with $\chi_s=0$ at $T=H=0$.
This quasiparticle contribution is also reflected in the Sommerfeld coefficient
$\gamma(H)$ measured by specific heat.
At low temperatures,
$\chi_s(H)/\chi_{\rm N}=\gamma(H)/\gamma_{\rm N}$.
In the spin-triplet case, by contrast, $\chi_s$ consists of both the intrinsic spin contribution
from the condensate and the quasiparticle contribution.

\begin{figure}
\begin{center}
\includegraphics[width=12cm]{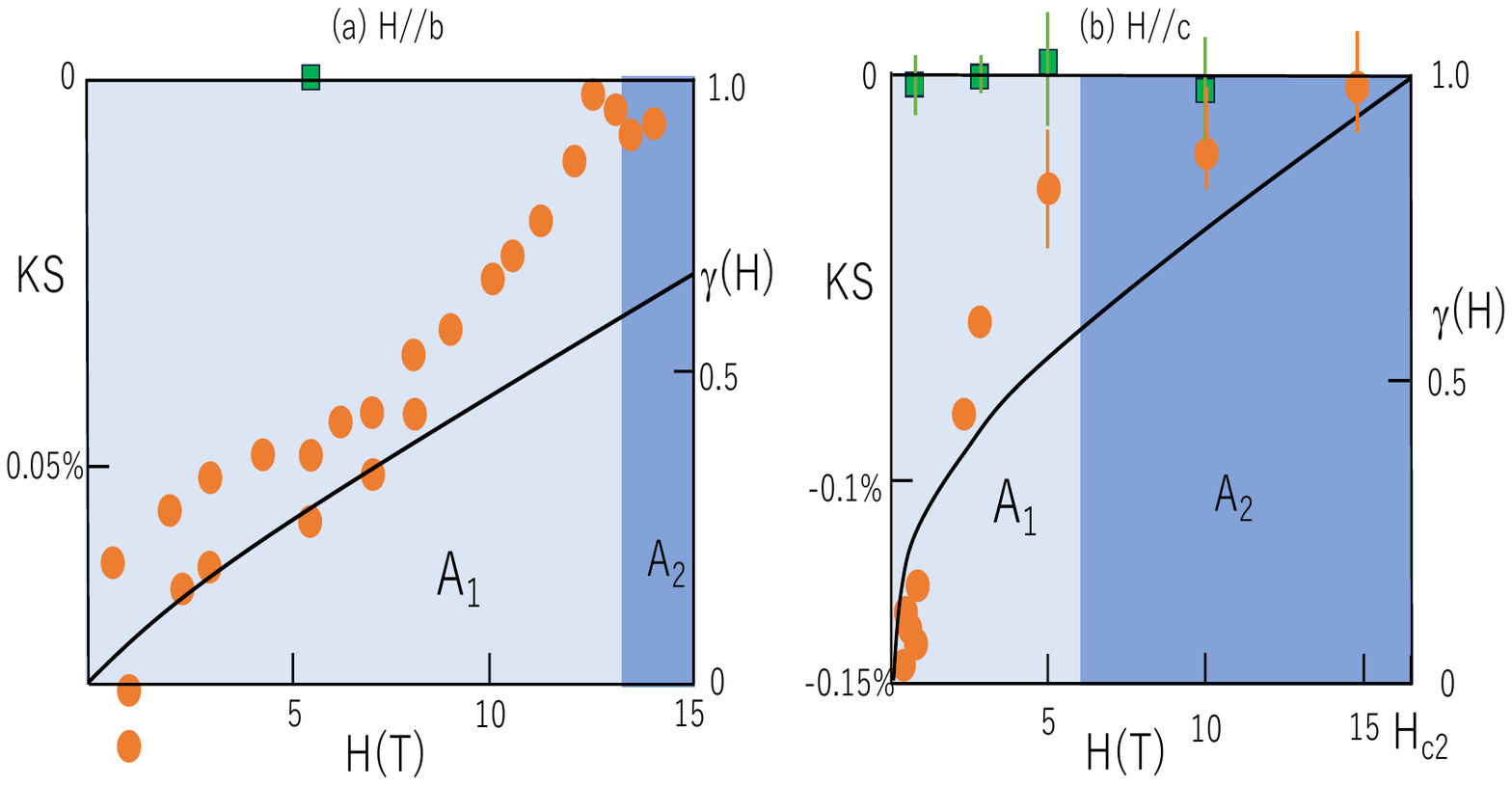}
\end{center}
\caption{
Field dependences of the Knight shift (KS)~\cite{matsumurachi}
(left scale) and the Sommerfeld coefficient
$\gamma(H)$~\cite{roman}, normalized by its normal-state value
(right scale).
(a) $H$$\parallel$$b$-axis.
The KS starts to deviate from $\gamma(H)$ around 8 T and reaches the
normal-state value at 14 T.
(b) $H$$\parallel$$c$-axis.
The KS reaches the normal-state value around 6 T.
These fields correspond to the lock-in fields of the d-vector.
The low- and high-field phases are denoted by $A_1$ and $A_2$, respectively.
\label{ksbc}}
\end{figure}

As shown in Figs.~\ref{ksbc}(a) and \ref{ksbc}(b), where we compare the KS data~\cite{matsumurachi}
with the $\gamma(H)$ data~\cite{roman}, the KS clearly deviates from $\gamma(H)$.
The deviation becomes pronounced around 8 T (a few tesla) for
$H\parallel b$ ($H\parallel c$), indicating that the d-vector rotation begins in these field ranges.
The spin susceptibility reaches the normal-state value, $\chi_{\rm N}$,
at 14 T for $H\parallel b$ and at 6 T for $H\parallel c$.
These fields correspond to the spin lock-in fields, above which
$\gamma(H)$ continues to increase whereas $\chi_s(H)$ remains equal to $\chi_{\rm N}$,
because the additional quasiparticles contribute only to the normal-state spin susceptibility.
We emphasize that these lock-in fields coincide precisely with the
$A_1$--$A_2$ transition fields shown in
Figs.~\ref{pd-abc}(b) and \ref{pd-abc}(c).

These gradual changes in the spin polarization $\mathbf{S}$ over a field range of several tesla
cannot be explained by a first-order phase transition between different pairing symmetries,
such as one arising from an accidental degeneracy between distinct irreducible representations.
Instead, they are naturally understood within the picture that the
spin-up $A_2$ phase becomes progressively more favorable than the spin-down
$A_1$ phase as the magnetic field increases.

As pointed out in Ref.~[\onlinecite{matsumurachi}], a kink in $H_{\rm c2}$
is observed at the lock-in field of 6 T for $H\parallel c$, as shown in
Fig.~\ref{pd-abc}(c).
This suggests the existence of a horizontal second-order transition line at 6 T,
consistent with our phase diagram in Fig.~\ref{pd-abc}(c).
In this connection, the kink observed in $\gamma(H)$ at 6 T~\cite{roman}
provides further support for our interpretation that this is a second-order phase transition.
Since $\gamma(H)$ is proportional to the entropy $S(H)$ at low temperatures,
the kink in $\gamma(H)$ implies a discontinuity in its derivative,
which, according to the Ehrenfest relation, is characteristic of a second-order phase transition.

\section{Discussion}

\subsection{Nodal structure and spin texture: Pairing symmetry and classification scheme}

Based on the arguments presented above, we now examine the possible pairing symmetry,
including both the orbital and spin degrees of freedom, within the framework of
$SO(3)^{\rm spin}\times D^{\rm orbital}_{\rm 2h}\times U(1)^{\rm gauge}$,
which corresponds to the finite spin-orbit coupling (SOC) scheme~\cite{machida0,annett}.
Among the one-dimensional odd-parity irreducible representations,
$^3A_{\rm 1u}$, $^3B_{\rm 1u}$, $^3B_{\rm 2u}$, and $^3B_{\rm 3u}$,
the $^3B_{\rm 3u}$ representation provides the most consistent interpretation
of the experimental results discussed above. More specifically,

\begin{eqnarray}
{
\Delta(k)=(\hat{\bf b}+i\hat{\bf c})\sin(ak_a/2).
}
\label{jumphi}
\end{eqnarray}

\noindent
This pairing state has line nodes only on the $\beta$ band,
while the $\alpha$ band remains fully gapped.
This gap structure is fully consistent with the angle-resolved specific-heat measurements~\cite{totsuka},
but inconsistent with the earlier results obtained using first-generation samples
with $T_{\rm c}=1.6$ K~\cite{kittaka}.

We make the following remarks.

\noindent
(1) The conventional identification of nodal structures based solely on power-law
temperature dependences observed in thermodynamic measurements,
such as thermal conductivity~\cite{metz,hayes,matsuda}
and penetration depth~\cite{shibauchi},
is difficult to apply to the present system.
Its multiband nature obscures the distinction among a full gap,
point nodes, and line nodes.

\noindent
(2) According to the Blount theorem~\cite{blount},
line nodes are not allowed in the strong-SOC limit,
where the spin and orbital degrees of freedom are tightly coupled.
Therefore, the identification of line nodes implies that the appropriate
group-theoretical classification should be based on the weak-SOC scheme.

\noindent
(3) In the identified pairing state,
$(\hat{\bf b}+i\hat{\bf c})\sin(ak_a/2)$ belonging to the
$^3B_{\rm 3u}$ representation,
the orbital component is unique.
This is consistent with the resonant ultrasound measurements~\cite{ramshaw},
which show no anomaly in the shear modes at the superconducting transition,
excluding the two orbital component scenario.

\subsection{Spontaneous magnetic moment below $T_{\rm c}$}

We point out a simple consequence of the GL theory developed in this work.
So far, we have treated the magnetic subsystem as being given \textit{a priori}.
Here we consider the effect of the superconducting order parameter
$\Delta_{\downarrow}$ on the magnetic subsystem.
The relevant coupling term is

\begin{eqnarray}
{
F_M=aM^2+\kappa M\Delta^2_{\downarrow}+\beta_MM^4,
}
\label{jumphi}
\end{eqnarray}

\noindent
where the GL coefficients satisfy $a>0$ and $\beta_M>0$.
Completing the square, we obtain

\begin{eqnarray}
{
F_M=a\Big(M+{\kappa\Delta^2_{\downarrow}\over 2a}\Big)^2
-{\kappa^2\Delta^4_{\downarrow}\over 4a}
+\beta_MM^4.
}
\label{jumphi}
\end{eqnarray}

\noindent
This expression immediately shows that a spontaneous magnetic moment

\[
M=-\frac{\kappa\Delta_{\downarrow}^2}{2a}
\]

\noindent
is induced below $T_{\rm c}$,
thereby lowering the total free energy through the bilinear coupling
between $M$ and $\Delta_{\downarrow}^2$.
The induced magnetic moment is directed antiparallel to the $a$-axis.

\subsection{Future studies}

We acknowledge that the present analysis is qualitative, or at best semi-quantitative,
and therefore requires further refinement to achieve a fully quantitative description.
This is particularly important for constructing the phase diagrams, which require
accurate parameter values such as $\kappa_i$.
In principle, these parameters can be obtained from electronic band-structure calculations,
since they originate from the particle-hole asymmetry about the Fermi level.
Moreover, the second transition temperature $T_{\rm c2}$, tentatively assigned to
0.1--0.3 K in the present work, should be determined experimentally.

Several important issues remain to be addressed.

\noindent
$\bullet$ We have predicted horizontal phase-transition lines and the associated
tetracritical point.
High-precision thermodynamic measurements are essential to verify these predictions.
In particular, the missing fourth internal transition lines for
$H$$\parallel$$a$ and $H$$\parallel$$c$ should be searched for experimentally.

\noindent
$\bullet$ Much remains to be explored under pressure.
Near the tetracritical pressure of $P=0.18$ GPa, an additional $A_0$ phase is predicted.
Its phase boundary should be determined experimentally, and the associated missing transition
line should be identified.
From a theoretical viewpoint, the critical behavior around this multicritical point
also deserves detailed investigation, since six second-order transition lines meet there.

\noindent
$\bullet$ As for the Cooper-pair symmetry, the line-node gap structure proposed in this work
should be examined experimentally.
Furthermore, because the present non-unitary state breaks time-reversal symmetry,
more sensitive $\mu$SR measurements should be performed to search for the associated spontaneous
internal magnetic field.
It would also be worthwhile to develop alternative experimental probes,
such as spin-current measurements.
The previous $\mu$SR experiments failed to detect such a signal~\cite{ajeesh,sonier1}.

\section{Conclusion and summary}

In this paper, we have further developed our theoretical understanding of the spin-triplet superconducting candidate UTe$_2$.
Our study was motivated by several recent experimental advances, including NMR measurements of the Knight shift and nuclear spin-lattice relaxation time $T_1$, as well as specific-heat measurements under ambient and applied pressure.
Taken together, these experiments provide compelling evidence for the existence of a second superconducting phase in the low-temperature and low-field region.
We have proposed a unified theoretical framework that consistently accounts for these observations and the corresponding phase diagrams in the $H$--$T$--$P$ space.

The principal new results presented in this paper, beyond those reported in our previous studies~\cite{machida1,machida2,machida3,machida4,machida5,machida6,machida7}, are summarized as follows.

\noindent
(1) The $H$--$T$ phase diagrams for magnetic fields applied along the principal $a$-, $b$-, and $c$-axes at ambient pressure all consist of two superconducting phases, $A_1$ and $A_2$.
Their overall structures are remarkably similar, each exhibiting a horizontal second-order transition line terminating at a tetracritical point.
We have identified the physical origin of these internal phase transitions as the field-induced transformation from the less stable spin-down $A_1$ phase to the more stable spin-up $A_2$ phase.

\noindent
(2) The hidden $A_2$ phase in the low-temperature and low-field region is now supported by several thermodynamic experiments.
We have clarified why this phase has remained elusive: its superfluid density is extremely small and is therefore masked by the dominant contribution from the $A_1$ phase.

These results ultimately point to a unique pairing symmetry in UTe$_2$,
\[
(\hat{\bf b}+i\hat{\bf c})\sin(ak_a/2),
\]
which belongs to the $^3B_{3u}$ irreducible representation within the weak spin-orbit coupling classification scheme.
This pairing state consistently explains the observed phase diagrams, the Knight-shift behavior, and the nodal gap structure inferred from thermodynamic experiments, thereby providing a unified description of the superconducting state in UTe$_2$.

\section*{Acknowledgements}
The author thanks S. Kittaka, Y. Shimizu, K. Totsuka, and Y. Tsutsumi for fruitful collaborations,
S. Kitagawa, K. Ishida, and H. Matsumura for sharing their NMR results and for valuable discussions;
and Y. Tokiwa, Sangyun Lee, and Xiaolong Liu for providing experimental information.
These experimental studies have strongly motivated the present work.
This work was supported by JSPS KAKENHI Grant No.~21K03455.

\end{document}